\documentclass[%
 reprint,
superscriptaddress,
 amsmath,amssymb,
 pra,
]{revtex4-2}
\usepackage{graphicx}
\usepackage[pdfpagemode=UseNone,pdfstartview=FitH,colorlinks=true,linkcolor=blue,citecolor=blue,urlcolor=blue]{hyperref}
\usepackage{placeins}
\usepackage[all]{hypcap}
\makeatletter
\let\saved@includegraphics\includegraphics
\AtBeginDocument{\let\includegraphics\saved@includegraphics}
\renewenvironment*{figure}{\@float{figure}}{\end@float}
\makeatother

\newcommand{\beginsupplement}{
  \setcounter{table}{0}  
  \renewcommand{\thetable}{S\arabic{table}} 
  \setcounter{figure}{0} 
  \renewcommand{\thefigure}{S\arabic{figure}}
  \setcounter{section}{0}
  \setcounter{equation}{0}
  \renewcommand{\theequation}{S\arabic{equation}}
}

\begin{document}
\title{Exponential suppression of bit or phase flip errors with repetitive error correction}
\author{Google Quantum AI}
\date{\today}

\email[Corresponding author (Z.~Chen):]{chenjimmy@google.com}
\email[\\Corresponding author (J.~Kelly):]{juliankelly@google.com}

\begin{abstract}

Realizing the potential of quantum computing will require achieving sufficiently low logical error rates \cite{preskill2018quantum}.
Many applications call for error rates in the  $10^{-15}$ regime \cite{shor1999polynomial, fowler2012surface, childs2018toward, campbell2019applying, kivlichan2020improved, gidney2019factor, lee2020efficient, lemieux2020resource}, but state-of-the-art quantum platforms typically have physical error rates near $10^{-3}$ \cite{ballance2016high, huang2019fidelity, rol2019fast, jurcevic2020demonstration, foxen2020demonstrating}. 
Quantum error correction (QEC) \cite{shor1995scheme, calderbank1996good, terhal2015quantum} promises to bridge this divide by distributing quantum logical information across many physical qubits so that errors can be detected and corrected.
Logical errors are then exponentially suppressed as the number of physical qubits grows, provided that the physical error rates are below a certain threshold.
QEC also requires that the errors are local and that performance is maintained over many rounds of error correction, two major outstanding experimental challenges. 
Here, we implement 1D repetition codes embedded in a 2D grid of superconducting qubits which demonstrate exponential suppression of bit or phase-flip errors, reducing logical error per round by more than $100\times$ when increasing the number of qubits from 5 to 21. 
Crucially, this error suppression is stable over 50 rounds of error correction.
We also introduce a method for analyzing error correlations with high precision, and characterize the locality of errors in a device performing QEC for the first time.
Finally, we perform error detection using a small 2D surface code logical qubit on the same device \cite{horsman2012surface, andersen2020repeated}, and show that the results from both 1D and 2D codes agree with numerical simulations using a simple depolarizing error model.
These findings demonstrate that superconducting qubits are on a viable path towards fault tolerant quantum computing.

\end{abstract}

\maketitle

\section{Introduction}

Many quantum error correction schemes can be classified as \textit{stabilizer codes} \cite{gottesman1997stabilizer}, where a single bit of quantum information is encoded in the joint state of many physical qubits, which we refer to as \textit{data qubits}.
Interspersed among the data qubits are \textit{measure qubits}, which periodically measure the parity of chosen combinations of data qubits.
These projective measurements turn undesired perturbations to the data qubit states into discrete errors which we track by looking for changes in the parity measurements.
The history of parity measurements can then be decoded to determine the most likely correction for such errors.
The error rate on the logical qubit is determined by the error rate on the physical qubits as well as the effectiveness of decoding.
If physical error rates are below a certain threshold determined by the decoder, then the probability of logical error per round of error correction  ($\epsilon_{\text{L}}$) should scale as:
\begin{equation}
\epsilon_{\text{L}} = C / \Lambda^{(d + 1) / 2} ,
\end{equation}
where $\Lambda$ is the exponential suppression factor, $C$ is a fitting constant, and $d$ is the code distance, which is related to the maximum number of physical errors allowed and increases with the number of physical qubits \cite{fowler2012surface, kelly2015state}.

Many previous experiments have demonstrated the principles of stabilizer measurements in various platforms such as NMR \cite{cory1998experimental, knill2001benchmarking}, ion traps \cite{moussa2011demonstration, nigg2014quantum, egan2020fault}, and superconducting qubits \cite{kelly2015state, takita2017experimental, wootton2020benchmarking, andersen2020repeated}.
However, achieving exponential error suppression in large systems is not a given, because typical error models for QEC do not include effects such as crosstalk errors. 
Moreover, exponential error suppression has never previously been demonstrated with cyclic stabilizer measurements, which are a key requirement for fault tolerant computing but put into play error mechanisms such as state leakage, heating, and data qubit decoherence during the measurement cycle \cite{kelly2015state, pino2020demonstration}.

In this work, we focus on two stabilizer codes. 
First, in the repetition code, qubits are laid out in a 1D chain which alternates between measure qubits and data qubits.
Each measure qubit checks the parity of its two neighbors, and all of the measure qubits check the same basis so that the logical qubit is protected from either $X$ or $Z$ errors, but not both. 
In the surface code \cite{bravyi1998quantum, fowler2012surface}, the qubits are laid out in a 2D grid which alternates between measure and data qubits in a checkerboard pattern.
The measure qubits further alternate between $X$ and $Z$ types, allowing for protection against both types of errors.
The repetition code will serve as a probe for exponential error suppression with number of qubits, while a small ($d=2$) primitive of the surface code will test the forward compatibility of our device with larger 2D codes.

\section{QEC with the Sycamore Processor}

\begin{figure}[t]
    \centering
    \includegraphics[width=89mm]{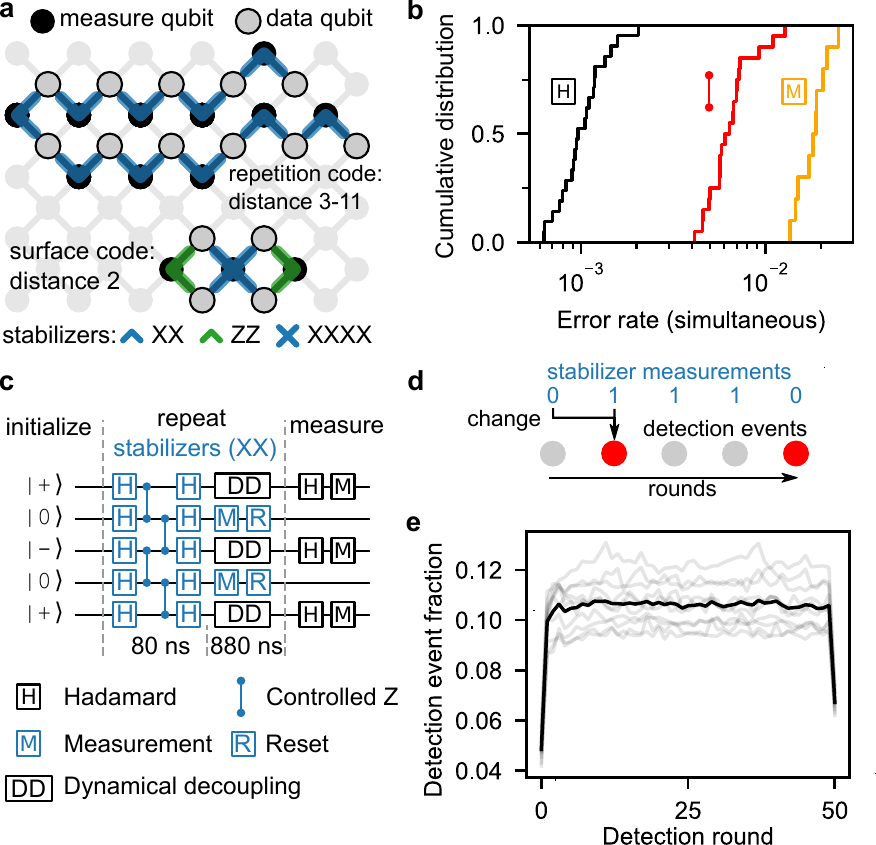}
    \caption{
    \textbf{Stabilizer circuits on Sycamore.} \textbf{a,} Layout of distance-11 repetition code and distance-2 surface code in the Sycamore architecture. In the experiment, the two codes use overlapping sets of qubits, which are offset in the figure for clarity.  \textbf{b,} Pauli error rates for gates and identification error rates for measurement. All benchmarks are for simultaneous operation. \textbf{c,} Circuit schematic for the phase flip code. Data qubits are randomly initialized into $|{+}\rangle$ or $|{-}\rangle$, followed by repeated application of $XX$ stabilizer measurements and finally $X$-basis measurements of the data qubits. \textbf{d,} Illustration of error detection events which occur when a measurement disagrees with the previous round. \textbf{e,} Fraction of measurements which detected an error versus measurement round for the $d=11$ phase flip code. The dark line is an average of the individual traces (gray lines) for each of the 10 measure qubits. The first (last) round also uses data qubit initialization (measurement) values to identify parity errors and generate detection events.}
    \label{fig:my_label}
\end{figure}

We implement QEC using a Sycamore processor \cite{arute2019quantum}, consisting of a 2D array of transmon qubits \cite{koch2007charge} where each qubit is tunably coupled to four nearest neighbors - the connectivity required for the surface code. 
Compared to Ref \cite{arute2019quantum}, this device has an improved design of the readout circuit, allowing for faster readout with less crosstalk and a factor of 2 reduction in readout error per qubit.
While this processor has 54 qubits like its predecessor, we used at most 21.
Figure 1a shows the layout of the $d=11$ (21 qubit) repetition code and $d=2$ (7 qubit) surface code in the Sycamore device, while Fig.\,1b summarizes the error rates of the components which make up the stabilizer circuits. 
Additionally, the typical coherence times for each qubit are $T_1=15\,\mu s$ and $T_2=19\,\mu s$. 

We note here two advancements in gate calibration. 
First, we use the reset protocol introduced in Ref.\,\cite{mcewen2020}, which removes population from excited states (including non-computational states) by sweeping the transmon past the readout resonator. 
This reset gate is appended after each measurement during QEC operation, and produces the ground state within 280\,ns with a typical error below 0.5\%. 
Second, we implement a 26\,ns controlled-Z gate using a direct swap between the states $|11\rangle$ and $|02\rangle$, similar to the gates described in \cite{foxen2020demonstrating, sung2020realization}. 
As in Ref.\,\cite{arute2019quantum}, the tunable qubit-qubit couplings allow these CZ gates to be executed with high parallelism, and up to 10 CZ gates are executed simultaneously for the 21 qubit repetition code. 
Using simultaneous cross-entropy benchmarking \cite{arute2019quantum}, we find that the median Pauli error for the CZ gates is 0.62\% (or an average error of 0.50\%).

We focused our repetition code experiments on the phase flip code where data qubits occupy superposition states and are sensitive to both energy relaxation and dephasing, making it more challenging to implement and more predictive of the performance of a surface code.
A 5-qubit unit of the phase flip code is shown in Fig.\,1c. 
This stabilizer circuit maps the $X$-basis parity of the data qubits onto the measure qubit, which is measured then reset, and this circuit is repeated in both space (across the 1D chain) and time. 
During measurement and reset, the data qubits are dynamically decoupled to protect the data qubits from various sources of dephasing \cite{supplement}. 
In a single shot of the experiment, we initialize the data qubits into a random string of $|+\rangle$ or $|-\rangle$ on each qubit. 
Then, we repeat stabilizer measurements across the chain over many rounds, and finally, we measure the state of the data qubits in the $X$ basis.

Our first pass at analyzing the experimental data is to turn measurements into \textit{error detection events}, which we find by comparing stabilizer measurements of the same measure qubit between adjacent measurement rounds. 
We refer to each possible spacetime location of a detection event (i.e. a specific measure qubit and measurement round) as a \textit{detection node}. 

In Fig.\,1e, for each detection node in a 50-round, 21-qubit phase flip code, we plot the fraction of experiments (76,000 total) where a detection event was observed on that node, or the \textit{detection event fraction}.
Overall, roughly 11\% of measurements signaled a detection event, except in the first and last round. 
At these two time boundary rounds, detections are determined by comparing the first (last) stabilizer measurement with data qubit initialization (measurement).
Importantly, the time boundary rounds are not subject to errors accumulated by the data qubits during measure qubit readout, illustrating the importance of running QEC for multiple rounds to accurately extract performance \cite{supplement}. 
Aside from these boundary effects, we find that the detection event fraction is stable across all 50 rounds of the experiment, a key finding for the feasibility of QEC. 
Previous experiments had observed rising detection event fractions \cite{kelly2015state}, and we attribute the stability of our system to our use of reset to remove leakage in every round \cite{mcewen2020}. 

\section{Correlations in error detection events}

\begin{figure*} [t]
    \centering
    \includegraphics[width=183mm]{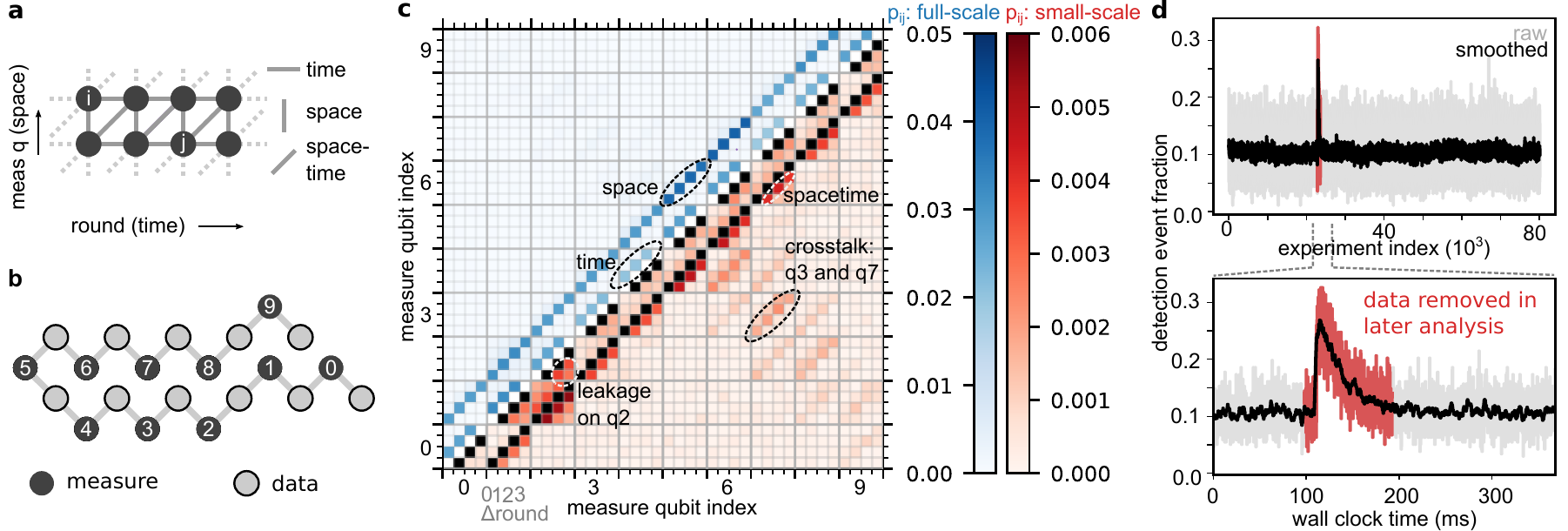}
    \caption{
\textbf{Analysis of error detections.} \textbf{a,} Detection event graph. Errors in the code trigger two detections (except at the ends of the chain), each represented by a node, and edges represent the expected correlations due to data qubit errors (spacelike and spacetimelike) and measure qubit errors (timelike) \textbf{b,} Ordering of the measure qubits in the repetition code. \textbf{c,} Measured two point correlations ($p_{ij}$) between detection events represented as a symmetric matrix. The axes correspond to possible locations of detection events, with major ticks marking measure qubits (space) and minor ticks marking difference in rounds (time). For the purposes of illustration, we have averaged together the matrices for 4-round segments of the 50-round experiment shown in Fig.\,1e, and also set $p_{ij} = 0$ if $i = j$.  The upper triangle shows the full scale, where only the expected spacelike and timelike correlations are apparent. The lower triangle shows a truncated color scale, highlighting unexpected correlations due to crosstalk and leakage. Note that crosstalk errors are still local in the 2D array. \textbf{d,} (Top) Observed high energy event in a time series of repetition code runs. (Bottom) Zoom in on high energy event, showing rapid rise and exponential decay of device wide correlated errors, and data which is removed when computing logical error probabilities.}
    \label{fig:my_label}
\end{figure*}

We next characterize the pairwise correlations between detection events. 
A Pauli error affecting any operation in the repetition code should produce exactly two detections (except at the spatial boundaries of the code) which come in three flavors \cite{kelly2015state}. 
First, an error on a data qubit usually produces a detection on the two neighboring measure qubits in the same round - a spacelike error. 
The exception is an error during the CZ gates, which may cause detection events offset by 1 unit in time and space - a spacetimelike error.
Finally, an error on a measure qubit which does not propagate to a data qubit will produce detections in two subsequent rounds - a timelike error. 
These rules are represented in the planar graph shown in Fig.\,2a, where expected correlations are drawn as graph edges between detection nodes.

We check how well Sycamore conforms to these expectations by computing the correlation probabilities between arbitrary pairs of detection nodes. 
Under the assumption that all correlations are pairwise and that error rates are sufficiently low, the probability of simultaneously triggering two detection nodes $i$ and $j$ can be estimated as
    \begin{align}
     p_{ij} \approx \frac{ \langle x_i x_j\rangle - \langle x_i \rangle\langle x_j \rangle }{(1 - 2\langle x_i\rangle)(1 - 2 \langle x_j\rangle)}\, ,
    \label{pij-approx}\end{align}
where $x_i = 1$ if there is a detection event and $x_i = 0$ otherwise, and $\langle x \rangle$ denotes an average over all experiments \cite{supplement}. 
The numerator can be understood as the covariance between detections in $i$ and $j$, while the denominator is an adjustment factor.
Note that $p_{ij}$ is symmetric between $i$ and $j$. 
In Fig.\,2c, we plot the correlation matrix for the data shown in Fig.\,1e. 
In the upper triangle, we show the full scale of the data, where the only visible correlations are either spacelike or timelike, demonstrating that error correlations in the device behave mostly as expected.

However, the sensitivity of this technique allows us to find features which do not fit the expected categories.
In the lower triangle, we plot the same data but with the scale truncated by nearly an order of magnitude.
The next most prominent correlations are spacetimelike, as we expect, but we also find two additional categories of correlations.
First, we observe detection correlations between non-adjacent measure qubits in the same measurement round.
While these non-adjacent qubits are far apart in the repetition code chain, they are in fact spatially close \cite{supplement} since the 1D chain is embedded in a 2D array, which suggests that while crosstalk exists in our system, it is short range.
Optimization of the frequencies in our system already mitigates crosstalk errors to a large extent \cite{klimov2020snake, supplement}, but further research is required to further suppress these errors. 
Second, we find excess correlations between measurement rounds that differ by more than 1.
We attribute these long lived correlations to the presence of leakage on the data qubits, which may be generated by a number of sources including gates \cite{rol2019fast}, measurement, and heating \cite{chen2016measuring, wood2018quantification}.
For the observed crosstalk and leakage errors, the excess correlations are around $3\times 10^{-3}$, an order of magnitude below the measured spacelike and timelike errors but well above the noise floor of the measurement of $2 \times 10^{-4}$. 

Having established that on average, the errors are mostly well-behaved, we now highlight a different kind of error correlation. 
In Fig.\,2d, we plot a time series of detection event fractions averaged over all measure qubits for each shot of an experiment. 
We clearly observe a sharp spike in the errors at a specific point in time, followed by an exponential decay. 
These types of events introduce significant correlated errors for roughly 0.5\% of all data taken \cite{supplement}, and we attribute them to high energy particles such as cosmic rays striking the quantum processor, also recently observed in Ref.\,\cite{vepsalainen2020impact}. 
For the purposes of understanding the typical behavior of our system, we remove data near these events (Fig.\,2d.), but note that these errors will need to be understood and mitigated \cite{karatsu2019mitigation, cardani2020reducing} for large-scale fault-tolerant computers. 

\section{Logical errors in the repetition code}

\begin{figure}[t]
    \centering
    \includegraphics[width=85mm]{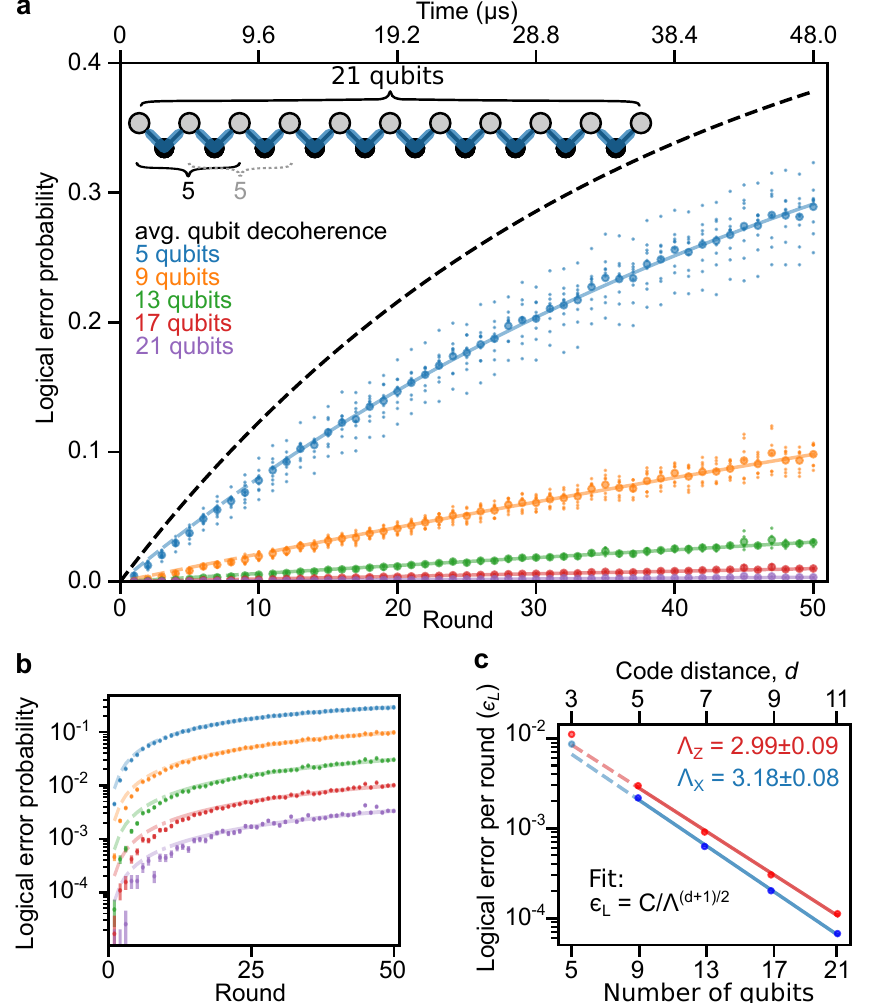}
    \caption{
    \textbf{Logical errors in the repetition code.} \textbf{a,} Logical error probability versus number of detection rounds and number of qubits for the phase flip code. Smaller code sizes are subsampled from the 21 qubit code as shown in the inset; small dots are data from subsamples and large dots are averages. \textbf{b,} Semilog plot of the averages from \textbf{a} showing even spacing in $\log\!{(\text{error probability})}$ between the code sizes. Error bars are estimated standard error from binomial sampling. The lines are exponential fits to data for rounds greater than 10. \textbf{c,} Logical error per round ($\epsilon_L$) vs. number of qubits, showing exponential suppression of error rate for both bit and phase flip, with extracted $\Lambda$ factors. The fit excludes $n_\text{qubits}=3$ to reduce the influence of spatial boundary effects \cite{supplement}.}
    \label{fig:my_label}
\end{figure}

We decode detection events and determine logical error probabilities following the procedure outlined in Ref.\,\cite{kelly2015state}. 
Briefly, we use a minimum weight perfect matching algorithm to determine which errors were most likely to have occurred given the observed detection events, and correct the final measured state of the data qubits in post-processing. 
A logical error occurs if the corrected final state is not equal to the initial state. 
We repeat the experiment and analysis while varying the number of detection rounds from 1 to 50 with a fixed number of qubits, 21. 
We determine logical performance of smaller code sizes by analyzing spatial subsets of the 21-qubit data, which reduces the amount of data required \cite{supplement}. 
These results are shown in Fig.\,3a, where we clearly observe a decrease in the logical error probability with increasing code size.
Figure 3b plots the same data on a semilog scale and illustrates the exponential nature of the error reduction.

To extract logical error per round ($\epsilon_L$), we fit the data for each number of qubits (averaged over spatial subsets) to $2 P_{\text{error}} = 1 - (1 - 2 \epsilon_L) ^ {n_{\text{rounds}}}$,  which expresses an exponential decay in logical fidelity with number of rounds. 
In Fig.\,3c, we show $\epsilon_L$ for the phase flip and bit flip codes versus qubit number. 
The data clearly demonstrates exponential suppression of logical errors, with more than $100 \times$ suppression in $\epsilon_L$ from 5 qubits ($\epsilon_L$ = $8.7\times10^{-3}$) to 21 qubits ($\epsilon_L$ = $6.7\times10^{-5}$). 
Additionally, we fit $\epsilon_L$ vs. code distance to Eqn.\,1 to extract $\Lambda$, which we plot in Fig.\,3c. 
We find $\Lambda_X=3.18 \pm 0.08$ for the phase flip code and $\Lambda_Z=2.99 \pm 0.09$ for the bit flip code \cite{supplement}. 

\section{Error budgeting and projecting QEC performance}

\begin{figure} [t]
    \centering
    \includegraphics[width=89mm]{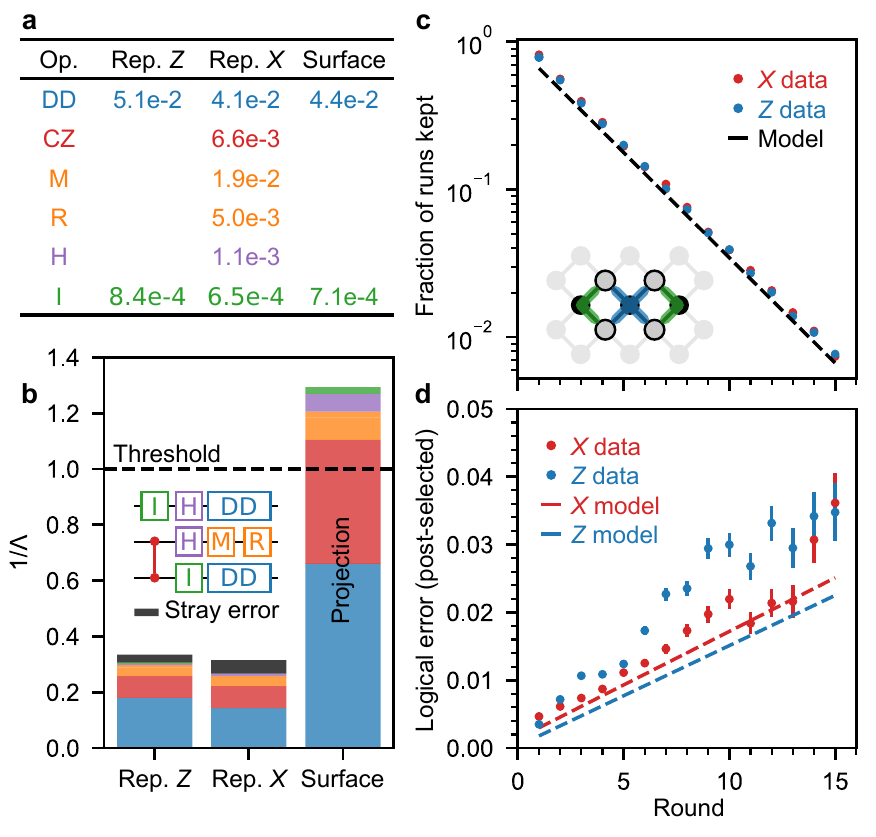}
    \caption{
\textbf{Error budgeting repetition and surface codes.}
\textbf{a,} Probability of depolarizing errors (bit flip errors for M and R) for various operations in the stabilizer circuit, derived from averaging quantities in Fig.\,1b. Note the idle gate (I) and dynamical decoupling (DD) values depend on the code being run because the data qubits occupy different states.
\textbf{b,} Estimated error budgets for the bit flip and phase flip codes, and projected error budget for the surface code, based on the depolarizing errors from \textbf{a}. The repetition code budgets slightly underestimate the experimental errors, and the discrepancy is labeled stray error. For the surface code, the estimated $1/\Lambda$ corresponds to difference in $\epsilon_L$ between a $d=3$ and $d=5$ surface code.
\textbf{c,} For the $d=2$ surface code, fraction of runs that had no detection events versus number of rounds, plotted with the prediction from a similar error model as the repetition code (dashed line). Inset: physical qubit layout of the $d=2$ surface code, 7 qubits embedded in a 2D array.
\textbf{d,} Surface code logical error probability among runs with no detection events versus number of rounds. Simulations from the same model as \textbf{c} (dashed lines) show good agreement. Error bars for \textbf{c} (not visible) and \textbf{d} are estimated standard error from binomial sampling with 240,000 experimental shots, minus the shots removed by post-selection in \textbf{d}.}
    \label{fig:my_label}
\end{figure}

To better understand our repetition code results and project surface code performance on the Sycamore architecture, we simulated our experiments using a depolarizing noise model, meaning that we inject a random Pauli error ($X$, $Y$, or $Z$) with some probability after each operation \cite{supplement}. 
The Pauli error probability for each type of operation is computed using averages of the data in Fig.\,1b and shown in Fig.\,4a. 
We perform two different types of simulations to compare our model to the  data. 
First, we run a direct simulation using the error rates in Fig.\,4a. to obtain a value of $\Lambda$ which should correspond to our measured values. 
Second, we simulate the experiment while individually sweeping operational error rates and observing how $1/\Lambda$ changes.
The relationship between $1/\Lambda$ and the component error rates is roughly linear \cite{supplement}, and the sensitivity coefficients obtained from the second simulation allow us to estimate how much each operation in the circuit increases $1/\Lambda$ (decreases $\Lambda$). 
The resulting error budgets for the phase and bit flip codes are shown in Fig.\,4b. 
Overall, measured values of $\Lambda$ are roughly 20\% lower than simulated values, which we attribute to mechanisms such as the leakage and crosstalk errors which are shown in Fig.\,2c and were not included in the simulations. 
Of the modeled contributions to $1/\Lambda$, the dominant sources of error are from the CZ gate and decoherence of the data qubits during measurement and reset.
In the same plot, we show the projected error budget for a surface code, where we find that overall performance must be improved to observe error suppression in a $d=5$ surface code compared to $d=3$.

Finally, we test our model against a distance-2 surface code logical qubit \cite{andersen2020repeated}.
We use seven qubits in the same Sycamore device to implement one weight-4 $X$ stabilizer and two weight-2 $Z$ stabilizers as depicted in Fig.\,1a. 
This encoding can detect any single error, but contains ambiguity in what correction corresponds to a given detection, so we discard any runs where we observe a detection event.
We show the fraction of runs where no errors were detected in Fig.\,4c for both logical $X$ and $Z$ preparations; we discard 27\% of runs each round, in good agreement with the model prediction. 
Logical errors can still occur after post-selection, for example with two simultaneous errors.
Following post-selection, we compute the logical error probability in the final measured state of the data qubits, shown in Fig.\,4d, where we find roughly $2 \times 10^{-3}$ error probability per round \cite{supplement}. 
The model slightly underestimates the logical error, with stray error similar to the repetition code case, giving us confidence that our surface code projections are accurate up to small corrections for crosstalk and leakage.

\section{Conclusion and outlook}
In this work, we show that a system with 21 superconducting qubits is stable when undergoing many repetitive stabilizer measurement cycles. 
By computing the probabilities of detection event pairs, we find that the physical errors detected on the device are localized in space and time to the $3 \times 10^{-3}$ level. 
Logical errors in the repetition code are exponentially suppressed when increasing the number of qubits from 5 to 21, even after 50 rounds of operation. 
Finally, we corroborate experimental results on both 1D and 2D codes with depolarizing model simulations and show that the Sycamore architecture is within a striking distance of the surface code threshold.

Nevertheless, many challenges remain on the path towards scalable quantum error correction. 
In the short term, our error budgets point to the salient research directions required to reach the surface code threshold: reducing the CZ gate error, and reducing data qubit errors during the measurement and reset cycle.  
Reaching this threshold will be an important milestone in quantum computing, but practical quantum computation will require $\Lambda \sim 10$ for the physical qubit overhead to be reasonable \cite{supplement}. 
Achieving this performance will require significant reductions in operational error rates, and maintaining a stable system over the course of a computation will require further research into mitigation of novel error mechanisms such as high energy particles.

\section{Author Contributions}
Z. Chen, K. Satzinger, H. Putterman, A. Fowler, A. Korotkov and J. Kelly designed the experiment. 
Z. Chen, K. Satzinger, and J. Kelly performed the experiment, and analyzed the data. 
C. Quintana, K. Satzinger, A. Petukhov, and Y. Chen developed the controlled-Z gate. 
M. McEwen, D. Kafri, A. Petukhov, and R. Barends developed the reset operation. 
M. McEwen and R. Barends performed experiments on leakage, reset, and high energy events in error correcting codes.
D. Sank and Z. Chen developed the readout operation.
A. Dunsworth, B. Burkett, S. Demura, and A. Megrant led the design and fabrication of the processor.
J. Atalya and A. Korotkov developed and performed the $p_{ij}$ analysis.
C. Jones developed the $1/\Lambda$  model and performed the simulations.
A. Fowler and C. Gidney wrote the decoder and interface software. 
S. Hong, K. Satzinger, and J. Kelly developed the dynamical decoupling protocols.
P. Klimov developed error mitigation techniques based on system frequency optimization.
Z. Chen, K. Satzinger, S. Hong, P. Klimov and J. Kelly developed error correction calibration techniques.
Z. Chen, K. Satzinger, and J. Kelly wrote the manuscript. 
S. Boixo, V. Smelyanskiy, Y. Chen, A. Megrant, and J. Kelly coordinated the team-wide error correction effort.
All authors contributed to revising the manuscript and writing the supplementary information. 
All authors contributed to the experimental and theoretical infrastructure to enable the experiment.

\section{Data availability}
The data that support the plots within this paper and other findings of this study are available from the corresponding authors upon reasonable request.

\onecolumngrid

\vspace{1em}
\begin{flushleft}
{\small Google Quantum AI}

\bigskip
{\small
\renewcommand{\author}[2]{#1$^\textrm{\scriptsize #2}$}
\renewcommand{\affiliation}[2]{$^\textrm{\scriptsize #1}$ #2 \\}

\newcommand{\xGoogle}{\affiliation{1}{Google Research}}

\newcommand{\xUMass}{\affiliation{2}{Department of Electrical and Computer Engineering, University of Massachusetts, Amherst, MA}}

\newcommand{\xPritzker}{\affiliation{3}{Pritzker School of Molecular Engineering, University of Chicago, Chicago, IL}}

\newcommand{\xUCR}{\affiliation{4}{Department of Electrical and Computer Engineering, University of California, Riverside, CA}}

\newcommand{\xUCSB}{\affiliation{5}{Department of Physics, University of California, Santa Barbara, CA}}

\newcommand{\xAWS}{\affiliation{$^\dagger$}{Present address: AWS Center for Quantum Computing, Pasadena, CA 91125, USA (Work was done prior to joining AWS)}}

\newcommand{\Google}{1}
\newcommand{\UMass}{2}
\newcommand{\Pritzker}{3}
\newcommand{\UCR}{4}
\newcommand{\UCSB}{5}
\newcommand{\AWS}{$\dagger$}

\author{Zijun Chen}{\Google},
\author{Kevin J.~Satzinger}{\Google},
\author{Juan Atalaya}{\Google},
\author{Alexander N.~Korotkov}{\Google,\! \UCR},
\author{Andrew Dunsworth}{\Google},
\author{Daniel Sank}{\Google},
\author{Chris Quintana}{\Google},
\author{Matt McEwen}{\Google,\! \UCSB},
\author{Rami Barends}{\Google},
\author{Paul V.~Klimov}{\Google},
\author{Sabrina Hong}{\Google},
\author{Cody Jones}{\Google},
\author{Andre Petukhov}{\Google},
\author{Dvir Kafri}{\Google},
\author{Sean Demura}{\Google},
\author{Brian Burkett}{\Google},
\author{Craig Gidney}{\Google},
\author{Austin G.~Fowler}{\Google},
\author{Harald Putterman}{\Google,\! \AWS},
\author{Igor Aleiner}{\Google},
\author{Frank Arute}{\Google},
\author{Kunal Arya}{\Google},
\author{Ryan Babbush}{\Google},
\author{Joseph C.~Bardin}{\Google,\! \UMass},
\author{Andreas Bengtsson}{\Google},
\author{Alexandre Bourassa}{\Google,\! \Pritzker},
\author{Michael Broughton}{\Google},
\author{Bob B.~Buckley}{\Google},
\author{David A.~Buell}{\Google},
\author{Nicholas Bushnell}{\Google},
\author{Benjamin Chiaro}{\Google},
\author{Roberto Collins}{\Google},
\author{William Courtney}{\Google},
\author{Alan R. Derk}{\Google},
\author{Daniel Eppens}{\Google}, 
\author{Catherine Erickson}{\Google},
\author{Edward Farhi}{\Google},
\author{Brooks Foxen}{\Google},
\author{Marissa Giustina}{\Google},
\author{Jonathan A.~Gross}{\Google},
\author{Matthew P.~Harrigan}{\Google},
\author{Sean D.~Harrington}{\Google},
\author{Jeremy Hilton}{\Google},
\author{Alan Ho}{\Google},
\author{Trent Huang}{\Google},
\author{William J. Huggins}{\Google},
\author{L.~B.~Ioffe}{\Google},
\author{Sergei V.~Isakov}{\Google},
\author{Evan Jeffrey}{\Google},
\author{Zhang Jiang}{\Google},
\author{Kostyantyn Kechedzhi}{\Google},
\author{Seon Kim}{\Google},
\author{Fedor Kostritsa}{\Google},
\author{David Landhuis}{\Google},
\author{Pavel Laptev}{\Google},
\author{Erik Lucero}{\Google},
\author{Orion Martin}{\Google},
\author{Jarrod R.~McClean}{\Google},
\author{Trevor McCourt}{\Google},
\author{Xiao Mi}{\Google},
\author{Kevin C.~Miao}{\Google},
\author{Masoud Mohseni}{\Google},
\author{Wojciech Mruczkiewicz}{\Google},
\author{Josh Mutus}{\Google},
\author{Ofer Naaman}{\Google},
\author{Matthew Neeley}{\Google},
\author{Charles Neill}{\Google},
\author{Michael Newman}{\Google},
\author{Murphy Yuezhen Niu}{\Google},
\author{Thomas E.~O'Brien}{\Google},
\author{Alex Opremcak}{\Google},
\author{Eric Ostby}{\Google},
\author{Bálint Pató}{\Google},
\author{Nicholas Redd}{\Google},
\author{Pedram Roushan}{\Google},
\author{Nicholas C.~Rubin}{\Google},
\author{Vladimir Shvarts}{\Google},
\author{Doug Strain}{\Google},
\author{Marco Szalay}{\Google},
\author{Matthew D.~Trevithick}{\Google},
\author{Benjamin Villalonga}{\Google},
\author{Theodore White}{\Google},
\author{Z.~Jamie Yao}{\Google},
\author{Ping Yeh}{\Google},
\author{Adam Zalcman}{\Google}
\author{Hartmut Neven}{\Google},
\author{Sergio Boixo}{\Google},
\author{Vadim Smelyanskiy}{\Google},
\author{Yu Chen}{\Google},
\author{Anthony Megrant}{\Google},
\author{Julian Kelly}{\Google}

\bigskip

\xGoogle
\xUMass
\xPritzker
\xUCR
\xUCSB
\xAWS

}
\end{flushleft}

\twocolumngrid

\bibliography{main}

\beginsupplement

\clearpage
\onecolumngrid
\begin{center}
	\textbf{\large Supplementary information for\\ 
``Exponential suppression of bit or phase flip errors with repetitive error correction"}
\end{center}
\bigskip
\twocolumngrid

\section{Data for bit flip code}
In addition to the phase flip code that is primarily described in the main text, we also ran a bit flip code for which the logical error rates are shown in Fig.\,3c of the main text. The experimental implementation of the bit flip code is similar to the phase flip code except for the following differences:

\begin{itemize}
    \item Initialization and measurements are performed in the $Z$ basis instead of $X$.
    \item The stabilizers used are $Z$ type instead of $X$ type, which means that the the data qubits do not have Hadamards at the beginning and end of each stabilizer round, and parity is measured in the $Z$ basis rather than $X$.
    \item We do not run dynamical decoupling pulses on the data qubits during measurement.
    \item Finally, prior to measurement in every round, we flip all of the data qubits with a $\pi$ pulse to ensure that the data qubits do not collapse into the ground state and remain there, which would artificially reduce logical error probabilities. 

\end{itemize}

\begin{figure}
\includegraphics[width=85mm]{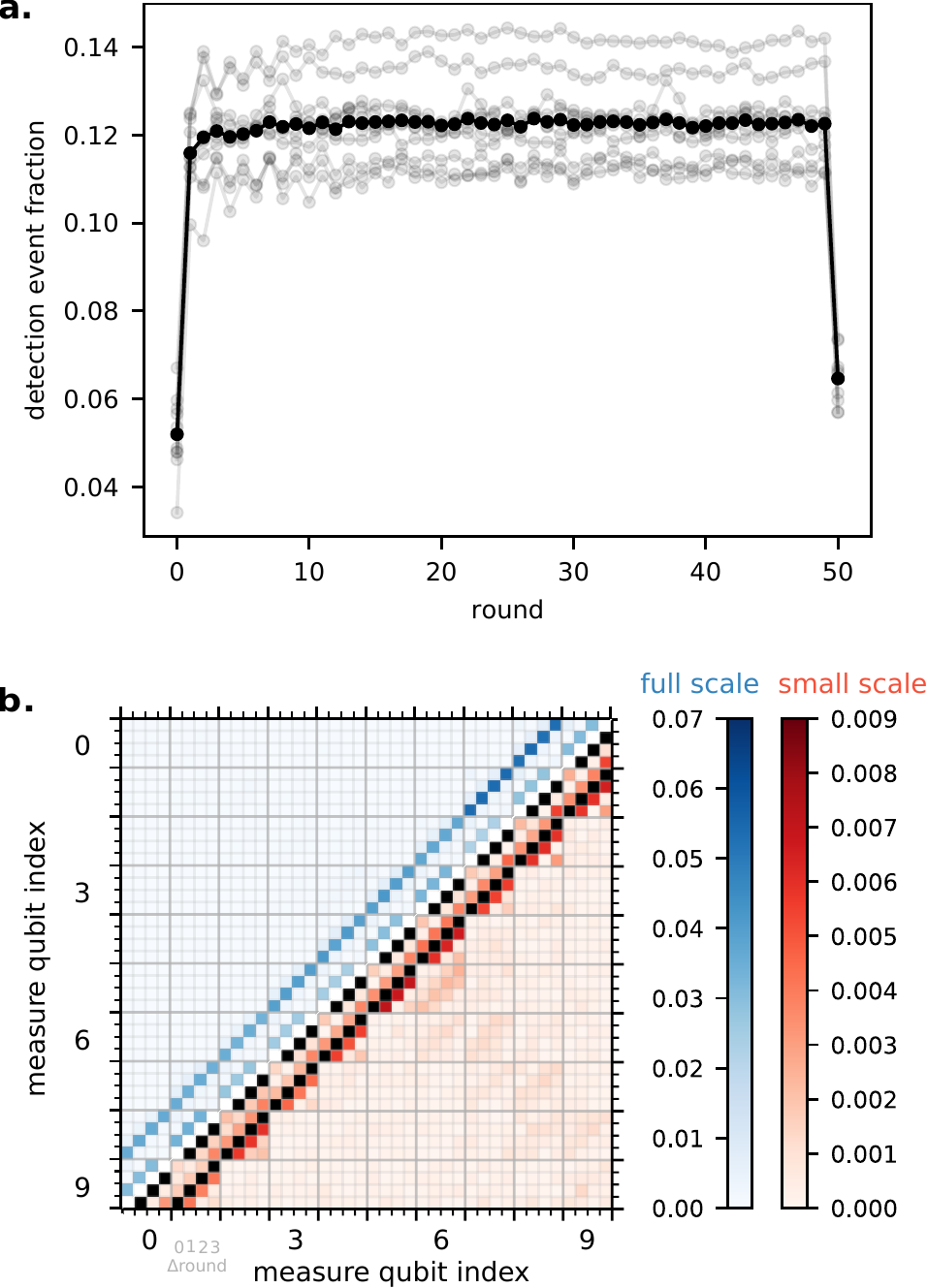}
\caption{\label{fig:bitflippij}  \textbf{a,} Detection event fraction for a 50 round bit flip code, similar to Fig.\,1d of the main text. \textbf{b,} $p_{ij}$ correlation matrix for the 50 round bit flip code, similar to Fig.\,2c of the main text}
\end{figure}

\begin{figure}
\includegraphics[width=85mm]{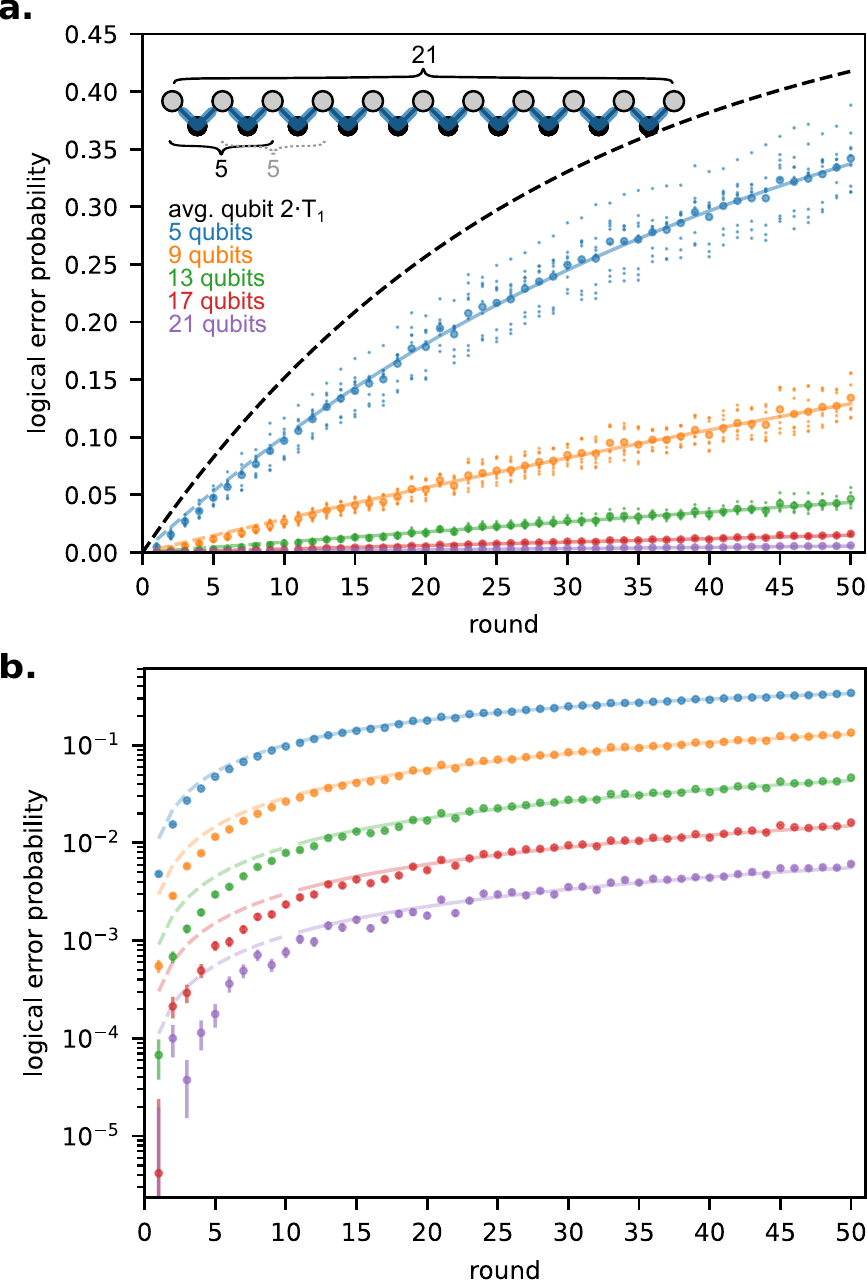}
\caption{\label{fig:bitfliplep} \textbf{a,} Logical error probabilities vs number of detection rounds for the bit flip code, similar to Fig.\,3a of the main text. \textbf{b,} Semilog plot of logical error probabilities, similar to Fig.\,3b of the main text. Lines depict fits to $2 P_{\text{error}} = 1 - (1 - 2 \epsilon_L) ^ {n_{\text{rounds}}}$ as in the main text for rounds greater than 10.}
\end{figure}

 In Fig.\,\ref{fig:bitflippij}, we show detection fractions and two point correlations for the 50 round bit flip code, and in Fig.\,\ref{fig:bitfliplep}, we show the logical error probabilities for rounds 1-50 of the bit flip code.
\section{Logical error probabilities without post-selection}

\begin{figure}
\includegraphics[width=85mm]{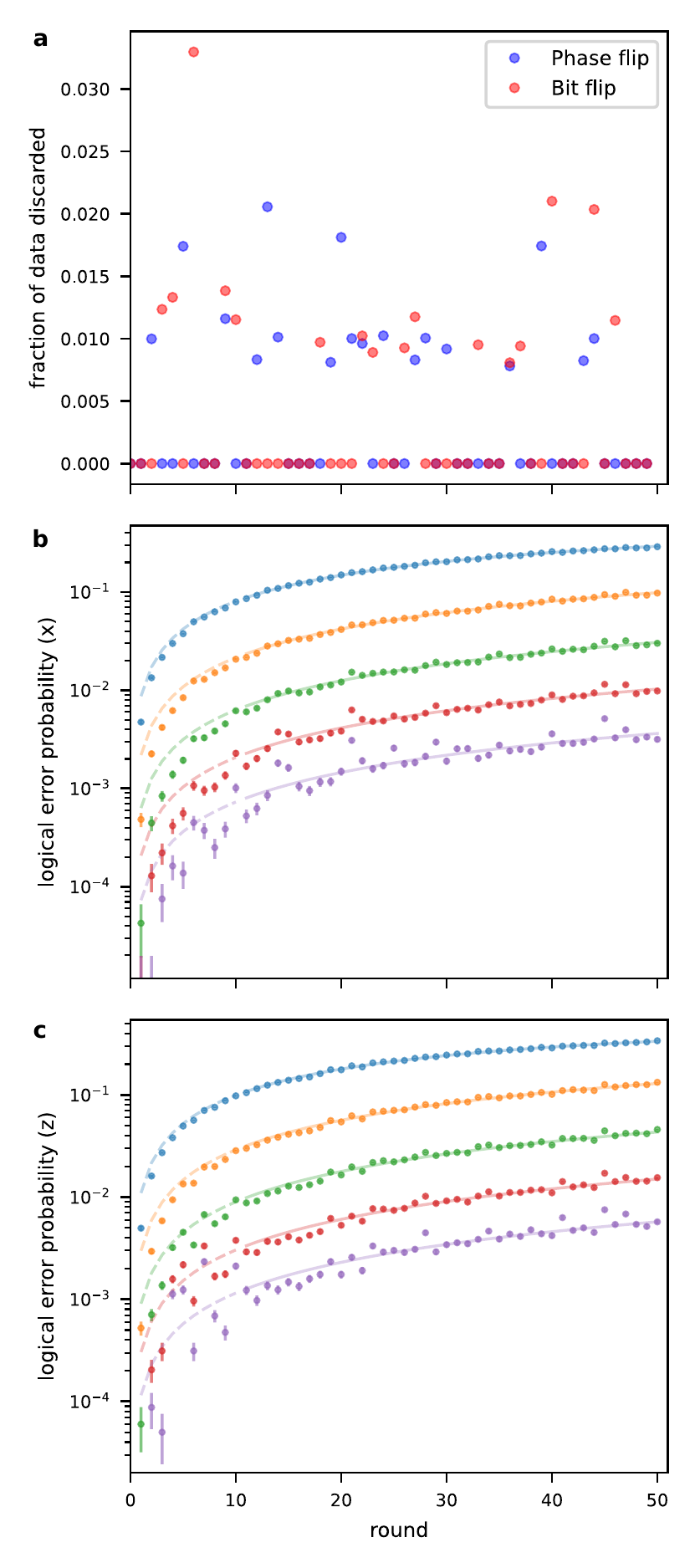}
\caption{\label{fig:nopostselect}  \textbf{a,} How much data was discarded for each run of the repetition code, in both $X$ and $Z$ bases \textbf{b.} Logical error probabilities for the phase flip code if high energy events are kept. Compare with Fig.\,3b. of the main text. \textbf{c,} Logical error probabilities for the bit flip code if high energy events are kept. Compare with Fig.\,\label{fig:bitfliplep}.}
\end{figure}

Logical error probabilities shown in Fig.\,3 of the main text were computed while excluding device-wide correlated error events which we attributed to high energy particles. 
In Fig.\,\ref{fig:nopostselect}, we show the fraction of data that was discarded for every number of rounds in the phase and bit flip codes, as well as the logical error probabilities. 
To within the uncertainty from fitting, values of $\Lambda_X$ and $\Lambda_Z$ do not change when we do not discard data. 
\section{The $d=2$ surface code}

\begin{figure}
\includegraphics{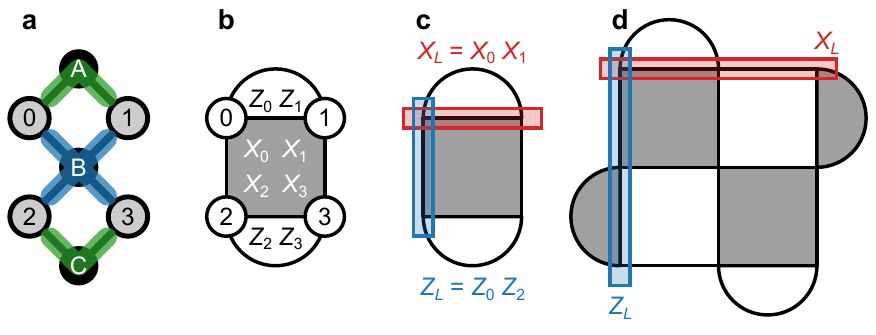}
\caption{
\textbf{Stabilizers and logical operators.}
\textbf{a,} Layout of the distance-2 logical qubit as depicted in Fig.~1a, with the data qubits labeled 0, 1, 2, 3, and the measure qubits labeled A, B, C.
\textbf{b,} The same logical qubit depicted in a more standard lattice surgery surface code notation, as in Ref.~\cite{fowler2019}. The $Z$ stabilizers are light tiles ($Z_0 Z_1$ and $Z_2 Z_3$), and the $X$ stabilizer is a dark tile ($X_0 X_1 X_2 X_3$).
\textbf{c,} The logical operators $X_L=X_0 X_1$ and $Z_L = Z_0 Z_2$, which cross at qubit 0, so $[X_L, Z_L] \neq 0$.
\textbf{d,} A distance-3 logical qubit and its logical operators, analogous to \textbf{c}, with 9 data qubits and 8 stabilizers.
}
\label{fig:d2-1}
\end{figure}

We implement a logical qubit in the distance-2 surface code, the smallest non-trivial example of a surface code logical qubit \cite{horsman2012, andersen2020}. The physical layout is depicted in Fig.~\ref{fig:d2-1}a-b, consisting of a $2 \times 2$ array of data qubits, indexed 0 to 3, subject to three stabilizer measurements $Z_0 Z_1$, $X_0 X_1 X_2 X_3$, and $Z_2 Z_3$.

Since there are only four data qubits, it is straightforward to write explicit quantum states for the $Z_L$ and $X_L$ eigenstates. Consider the case where the three stabilizer values are all +1. Then, the logical qubit exists in the two-dimensional ground state manifold of the Hamiltonian \cite{kitaev2003}
\begin{equation}
H = - X_0 X_1 X_2 X_3 - Z_0 Z_1 - Z_2 Z_3.
\end{equation}
We can isolate specific logical states using the logical operators $Z_L=Z_0 Z_2$ and $X_L=X_0 X_1$ shown in Fig.~\ref{fig:d2-1}c. For example, $|0_L\rangle$ (+1 eigenstate of $Z_L$) is the unique ground state of $H - Z_L$. An alternative way to identify $|0_L\rangle$ is to start with $|\psi_0\psi_1\psi_2\psi_3\rangle = |0000\rangle$, which is a +1 eigenstate of $Z_L$ and both $Z$ stabilizers, and then project it into the $X_0 X_1 X_2 X_3 = +1$ subspace with the projection operator $(1 + X_0 X_1 X_2 X_3)/2$. The logical states are
\begin{align*}
|0_L\rangle &= (|0000\rangle+|1111\rangle)/\sqrt{2} \\
|1_L\rangle &= X_L|0_L\rangle = (|0011\rangle+|1100\rangle)/\sqrt{2} \\
|+_L\rangle &= (|0_L\rangle+|1_L\rangle)/\sqrt{2} \\ 
            &= (|0000\rangle+|1111\rangle+|0011\rangle+|1100\rangle)/\sqrt{4} \\
|-_L\rangle &= (|0_L\rangle-|1_L\rangle)/\sqrt{2} \\
            &= (|0000\rangle+|1111\rangle-|0011\rangle-|1100\rangle)/\sqrt{4}.
\end{align*}

It is also possible for some stabilizer values to be $-1$. For example, if $X_0 X_1 X_2 X_3 = -1$ but the others are $+1$, then we identify $|0_L\rangle = (|0000\rangle-|1111\rangle)/\sqrt{2}$, differing from the $+1$ case by $Z_0$ (or any $Z_i$). Initializing to $|0000\rangle$ and projectively measuring $X_0 X_1 X_2 X_3$, this would be the outcome half the time (also see Fig.~\ref{fig:d2-3}a).

\begin{figure}
\includegraphics{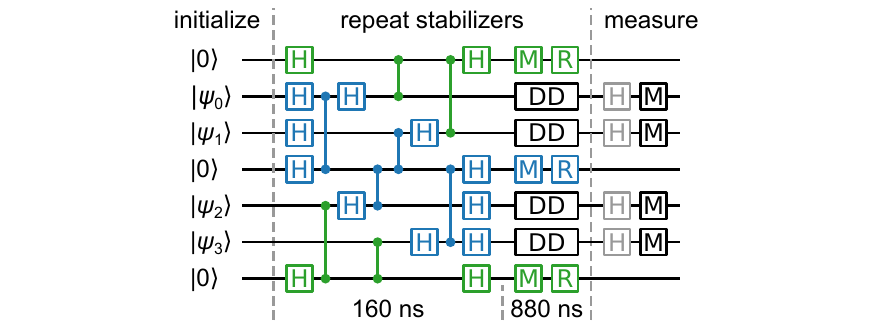}
\caption{
\textbf{Surface code quantum circuit.}
Quantum circuit implementing repeated $Z$ (green) and $X$ (blue) stabilizers, analogous to Fig.~1c. The stabilizer circuit is longer (four CZ layers) because of the weight-4 $X$ stabilizer. For $X_L$ logical measurements, we include Hadamard gates on each data qubit prior to measurement, shown in gray; these are omitted for $Z_L$ logical measurements.
}
\label{fig:d2-2}
\end{figure}

\begin{figure}
\includegraphics{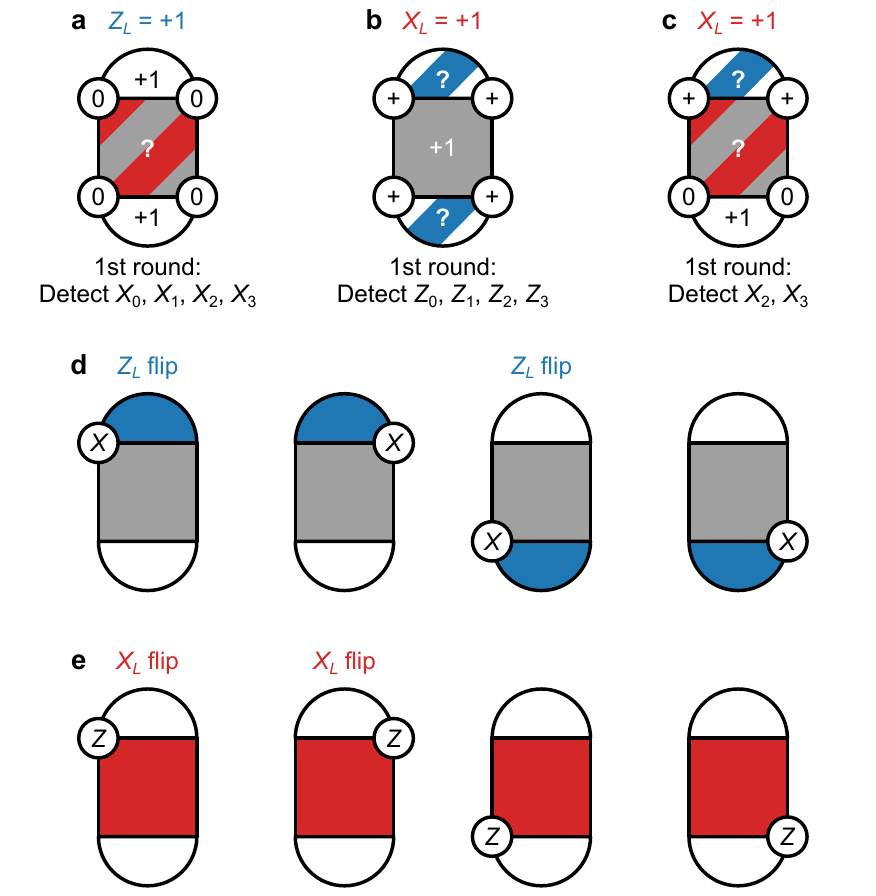}
\caption{
\textbf{Error detection.}
\textbf{a,} Example initialization to $|0000\rangle$ prior to the first round of stabilizer measurements. This is a $+1$ eigenstate of $Z_L$ and both $Z$ stabilizers. In the first round, any $X$ error can be detected. However, the first $X$ stabilizer measurement will be random, so no $Z$ errors can be detected.
\textbf{b,} $|{+}{+}{+}{+}\rangle$ is a $+1$ eigenstate of $X_L$ and the $X$ stabilizer. In the first round, any $Z$ error can be detected, but the two $Z$ stabilizers will have random values.
\textbf{c,} $|{+}{+}00\rangle$ is a $+1$ eigenstate of $X_L$ and the lower $Z$ stabilizer. As in \textbf{a}, the first $X$ stabilizer measurement will be random, so no $Z$ errors can be detected, risking a logical error $X_L=-1$.
\textbf{d,} Illustration of the detected syndrome for one $X$ error. Note $X_0$ and $X_1$ have the same syndrome, but $X_0$ flips $Z_L$ while $X_1$ does not. $X_2$ and $X_3$ are similar.
\textbf{e,} Illustration of the detected syndrome for one $Z$ error. All four have the same syndrome, but $Z_0$ and $Z_1$ flip $X_L$ while $Z_2$ and $Z_3$ do not.
In \textbf{d}-\textbf{e}, there is an implicit decoding procedure: for flipped $X_0X_1X_2X_3$, insert $Z_0$ correction; for flipped $Z_0Z_1$, insert $X_0$ correction; and for flipped $Z_2Z_3$, insert $X_2$ correction. When this correction is the wrong choice, which happens for about half of error events, we get logical errors.
}
\label{fig:d2-3}
\end{figure}

In our experiments, we explore all 8 stabilizer value combinations, which is representative of stabilizer values that would be encountered by a long-lived logical qubit. In particular, we initialize the data qubits to each of the 16 possible bitstrings, such as $|0111\rangle$. For experiments in the logical $Z$ basis, we proceed directly with stabilizer measurements, and the $Z$ stabilizers and $Z_L$ are already well-defined (for $|0111\rangle$, $Z_0 Z_1 = -1$, $Z_2 Z_3 = +1$, and $Z_L = -1$). The first $X_0 X_1 X_2 X_3$ measurement is randomly $\pm 1$. For experiments in the logical $X$ basis, we perform Hadamards on all four data qubits before proceeding with the stabilizer measurements, so $|0111\rangle$ becomes $|{+}{-}{-}{-}\rangle$. Now the $X$ stabilizer and $X_L$ are well-defined (for $|{+}{-}{-}{-}\rangle$, $X_0 X_1 X_2 X_3 = -1$ and $X_L = -1$), and the first $Z$ stabilizer measurements are each randomly $\pm 1$. We show the specific quantum circuit for these experiments, analogous to Fig. 1c, in Fig.~\ref{fig:d2-2}.

Note that to prepare a logical $X_L$ or $Z_L$ eigenstate, it is important to initialize all the data qubits in the same basis ($X$ or $Z$) as the intended logical qubit state. Then, the data qubit state is an eigenstate of all the stabilizers of the same type as the logical operator, and any errors of the opposite type can be detected in the first round. We show standard $Z$ and $X$ initializations in Fig.~\ref{fig:d2-3}a-b. Alternatively, consider $|{+}{+}00\rangle$, shown in Fig.~\ref{fig:d2-3}c, which is employed in Ref.~\cite{andersen2020}. The first $X_0 X_1 X_2 X_3$ measurement will be random, so no $Z$ errors can be detected on the first round, risking a logical error in $X_L$. Moreover, although $|{+}{+}00\rangle$ is an eigenstate of $X_L = X_0 X_1$, it is not an eigenstate of $X_L^\prime = (X_0 X_1 X_2 X_3)X_L = X_2 X_3$, an equally valid logical operator.

This encoding can detect any single error, but because it is only distance-2, the code cannot be used to correct for errors, as shown in Fig.~\ref{fig:d2-3}d-e. Any single error on a data qubit leads to an ambiguous syndrome, where it is unclear if a logical operator has been affected. This is distinct from the larger distance-3 logical qubit (see Fig.~\ref{fig:d2-1}d), where any single error can be corrected unambiguously (distance-$d$ can accommodate any $(d-1)/2$ errors).

Consequently, any time we observe a detection event in a run, we simply discard that run. As we increase the number of rounds, we increase the probability that there has been a detection event, so the fraction of runs we keep decreases exponentially, as shown in Fig.\,4c of the main text. Empirically, we remove about 27\% of runs each round, which agrees well with simulations of the experiment. 

At the end of each run, we measure the data qubits in the basis matching the logical basis of the experiment, either $X$ or $Z$, and evaluate the appropriate logical operator. We identify a logical error if the logical measurement outcome differs from the value we initialized. By post-selecting only runs without detection events, we avoid most logical errors. However, two simultaneous errors can be undetectable and lead to logical errors, such as $X_0 X_1$, which flips $Z_L$. Following post-selection, the probability of a logical error is about 0.002 each round, as shown in Fig. 4d. Specifically, for $X$ basis, we observe $0.0016\pm 0.0001$ error per round, and for $Z$ basis, $0.0027\pm 0.0001$ (linear fit uncertainties). For comparison, in Ref.~\cite{andersen2020}, about 60\% of runs are removed each round, and the logical error probability is about 0.03 each round.

\begin{figure}
\includegraphics{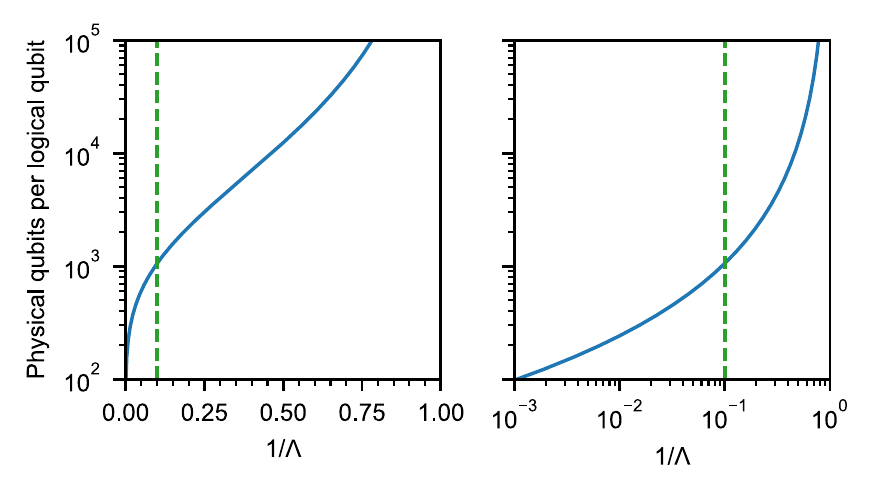}
\caption{
\textbf{Physical qubits per logical qubit.}
We estimate the physical qubits required for one logical qubit to achieve an overall logical error suppression of $10^{-12}$ as a function of the inverse error suppression factor $1/\Lambda$, marking $\Lambda=10$ with a vertical line. Left: semi-log, right: log-log.
}
\label{fig:d2-4}
\end{figure}

In Fig 4b, we project the error suppression factor $\Lambda$ for the surface code. Modest performance improvements will be needed to achieve $\Lambda > 1$, which would be a clear demonstration of operating below threshold error rates, where making the code larger makes it better (even if the absolute error rate is worse than a physical qubit). However, a practical surface code quantum computer would benefit from $\Lambda \sim 10$, which vastly decreases the required physical qubits per logical qubit for a given logical error rate. For example, suppose we want an overall logical error suppression $1/\Lambda^{(d+1)/2} = 10^{-12}$ for a practical computation. For a given $\Lambda$, we can solve for distance $d$ and estimate the required number of physical qubits per logical qubit as roughly $2d^2$, as shown in Fig.~\ref{fig:d2-4}. For $\Lambda=10$, this corresponds to roughly 1000 physical qubits (distance-23).

\FloatBarrier 
\section{Quantifying Lambda}
Accurately benchmarking the performance of quantum error correction can be confounded by artifacts if experiments are not carefully designed. 
In particular, boundary effects can introduce different error characteristics that must be understood. 
Here, we study two types of boundary effects. 
The first is qubits at code boundaries, which interact with a reduced number of stabilizers and thus participate in a reduced number of entangling gates and may decrease the number of physical errors present.
Second, data qubits are subject to less errors in the first round of the code than in the steady-state, and data qubit measurement errors are only relevant in the final round of measurements. 

\begin{figure}
\includegraphics{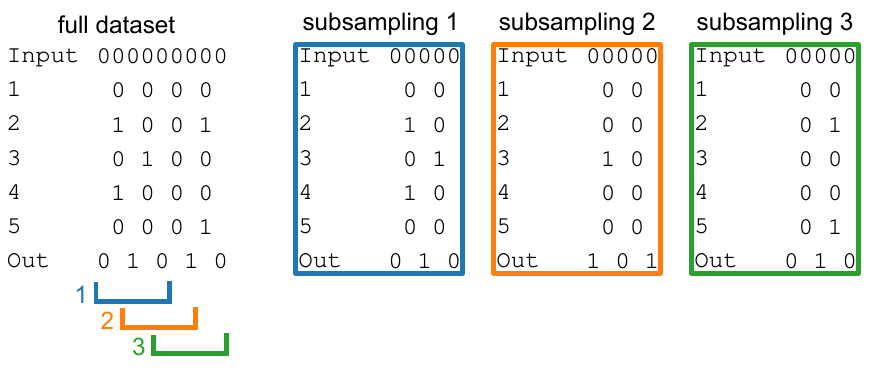}
\caption{
Example of subsampling a $d=5$ repetition code dataset into 3 $d=3$ repetition code datasets. 
}
\label{fig:quantlambda-1}
\end{figure}
In our analysis of the repetition code, we use the technique of subsampling outlined in the supplementary materials of \cite{kelly2015state}. 
In order to, for example, compare the performance of a d=11 repetition code to a d=3 repetition code, we take a single dataset for the d=11 code, perform matching analysis, then subsample this dataset into a collection of d=3 datasets and perform matching analysis on each sub-dataset. 
Generally, a repetition code of distance $d_s$ can be subsampled from a larger code of distance $d$, where $n=d - d_s + 1$ is the number of unique datasets one could produce. 
This can be understood by considering a line of 9 qubits (for $d=5$), and uniquely choosing a line of 5 qubits (for $d=3$) along it, as shown in Fig.\,~\ref{fig:quantlambda-1}.

Subsampling has a number of practical advantages. 
First and foremost, the experimental burden of acquiring data is reduced. 
In order to quantify the performance of a distance $d$ repetition code as well as all possible configurations of smaller code distances, without subsampling we would need to perform $n_\text{experiments}=\sum_{n=1}^{(d-1)/2,\,\text{odd}} d - 2n$. 
In the case of $d=11$, subsampling reduces the datasets needed by a factor of 25. 
Additionally, by using only a single source dataset, we enforce self-consistency in error rates between code distances and reduce sensitivity to systematic errors and system drift that may occur between data acquisition runs. 
Alternatively, one could collect only a single dataset for each code distance. 
However, qubits typically have performance variations and the choice of which qubits for which code distance at what time will introduce bias or noise into benchmarking. 

\begin{table}[]
\caption{
Pauli error rates (bit flip error rates for measurement and reset) used in subsequent simulations.
}
\begin{ruledtabular}
\begin{tabular}{ll}
{Operation} & {Error rate} \\ \hline
H                  & 1e-3           \\
CZ                 & 5e-3           \\
M                  & 2e-3           \\
R                  & 5e-3           \\
Idle (M + R)       & 4.4e-2         \\
Idle (H)           & 7e-4          
\end{tabular}
\end{ruledtabular}
\label{table:simerrorrates}
\end{table}

In order to understand boundary effects and their impact on repetition code data, we perform simulations using an uncorrelated depolarized Pauli error model. 
Here, we use a simple error model described by Table \ref{table:simerrorrates}, where every qubit shares identical error rates. 
Given these probabilities, we simulate 100,000 runs of a 21 qubit repetition code over 10 QEC rounds. 

\begin{figure}
\includegraphics{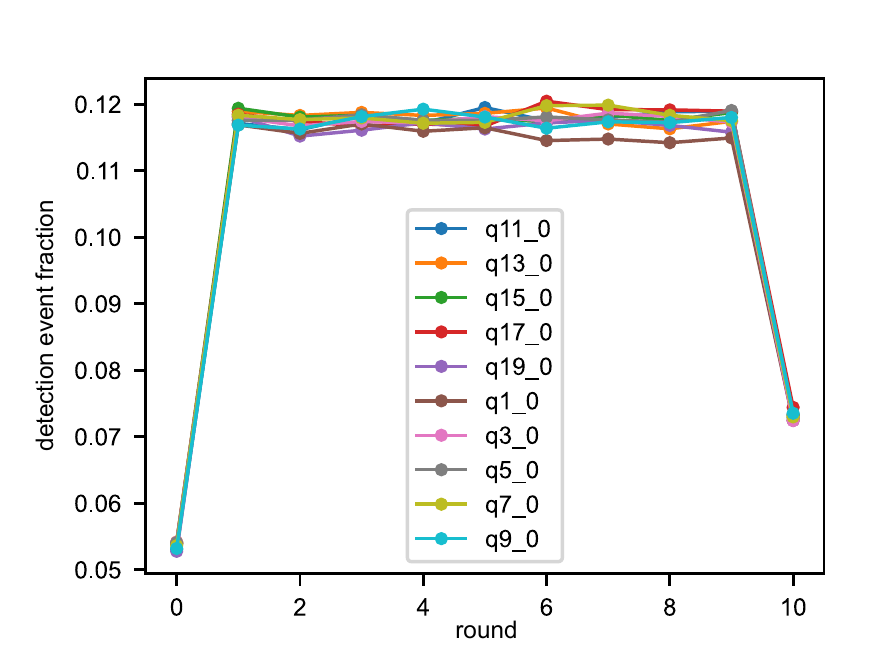}
\caption{
Simulated repetition code data for 10 QEC rounds and 21 qubits. The plot shows detection event fraction as a function of round. We find a uniform behavior of detection fraction in the intermediate rounds, and different values at the first and last rounds of the code, which differ in circuit structure.
}
\label{fig:quantlambda-2}
\end{figure}

We process this simulated data to explore the detection event fraction as a function of round, per qubit. 
We find that the first and last round deviate from the steady-state detection event round, as seen in Fig\,\ref{fig:quantlambda-2}. 
This discrepancy comes from a difference in circuit structure as well as initial conditions. 
Before initialization, all qubits begin in the $|0\rangle$ state and suffer no Idling error during the $M + R$ operations that subsequent rounds do. 
In the last round, the stabilizer outcomes are determined from the final data qubit measurements, and require no data qubit idling or entangling gates. 
These differences manifest in smaller error rates and thus smaller detection event fractions associated with these rounds. 

\begin{figure}
\includegraphics{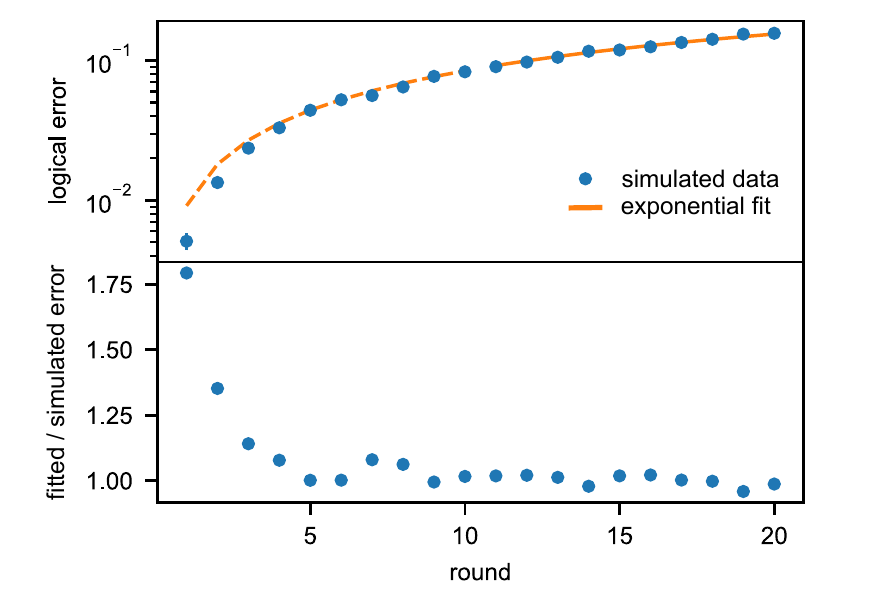}
\caption{
Logical error probabilities fitted to an exponential model of logical error rate (dashed, solid lines) (distance-3, repetitions = 10,000). At low rounds, we see deviations from the exponential fit due to boundary effects at the start and end of the code, where error rates are reduced as compared to the steady-state of the experiment. This can be seen in the lower graph, where we plot fitted error over simulated error. At low rounds, we find up to nearly a factor of 2 discrepancy. To mitigate this effect, we fit the exponential only to rounds greater than 10. Similar fits can be seen in Fig.\,3 of the main text, and in Fig.\,\ref{fig:bitfliplep} and Fig.\,\ref{fig:nopostselect}.
}
\label{fig:quantlambda-4}
\end{figure}

This non-uniformity in detection event fraction must be accounted for when analyzing $\Lambda$. 
In benchmarking QEC, we seek to quantify the logical error rate in the steady-state, but these boundary effects indicate the error rate is slightly different at the beginning and end of the code. Due to this effect, the logical error probabilities will deviate slightly from an exponential decay. 
To mitigate this behavior, we choose to fit an exponential decay to only experiments with a large number of rounds (greater than 10), where this effect is minimized. 
This can be seen in Fig.\,\ref{fig:quantlambda-4}, where in this simple model we see logical error probabilities that deviate from an exponential model (dashed, solid lines) at small numbers of rounds. 
In this regime, the logical error probabilities outperform the steady state and are not predictive of future QEC performance. 
This discrepancy, here up to a factor of 2, can vary depending on circuit construction and hardware. 

\begin{figure}
\includegraphics{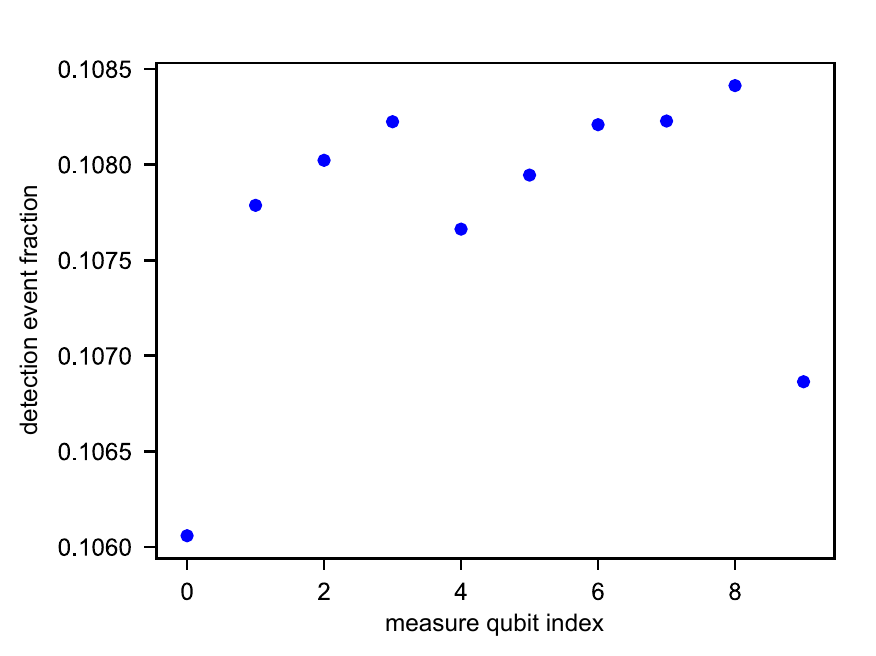}
\caption{
Detection event fraction vs measure qubit index for a 21 qubit repetition code. Detection event fraction for measure qubits at the edge of the code (index = 0, 9) have lower detection event fraction, as data qubits on the boundary participate in fewer entangling gates.
}
\label{fig:quantlambda-3}
\end{figure}

In addition to time boundary effects, spatial boundary effects also exist for qubits located at the edge of the code, which participate in less entangling gates. 
This can be seen in Fig.~\ref{fig:quantlambda-3}, where the measure qubits at the edges of a simulated 21 qubit repetition code have lower detection event fraction. 
This introduces a small but systematic difference in comparing subsampled data to experiments that are run in isolation. 

\FloatBarrier  
\section{Circuit simulations with Pauli noise}

This section describes simulations that approximate errors in the experiment as Pauli errors sampled from probability distributions and inserted into a circuit of Clifford gates. 
In many quantum error-correcting codes, including repetition codes and surface codes, the bulk of the encoded operations consist only of gates from the Clifford group~\cite{Got1997}; the exception is the need to enact logical non-Clifford gates, such as through magic-state distillation~\cite{BK2005}, which is needed in a fault-tolerant quantum computer but beyond the scope of logical memory experiments like this work.  
A circuit composed entirely of Clifford gates can be simulated efficiently using the Gottesman-Knill theorem~\cite{AB2006}, and this description includes noisy circuits where the noise is a probability distribution for randomly inserting a Pauli operator after each gate.  
Moreover, for stabilizer codes~\cite{Got1997}, the stabilizers are Pauli operators which can be measured by Clifford gates, so it is convenient to represent errors as a distribution of Pauli errors. 
We employ this model here --- Clifford circuits with Pauli errors --- because the simulations can easily scale to modeling large surface codes, such as a distance-23 surface code requiring at least 1057 qubits.

We employ circuit simulations to attempt to understand the relative contributions of errors from different operations, also known as error budgeting.  
This proceeds in two stages.  
First, we run simulations of the repetition codes with circuit-noise parameters informed by benchmarking component operations, such as CZ gate error from cross-entropy benchmarking and idling qubit error from measuring $T_1$ and $T_2$.  
We compare the logical error rate in the simulations with the logical errors in the experiment, and see close agreement. 
We also discuss possible explanations for the gap between experiment and simulation.

Second, we use simulations to estimate the relative contributions of component errors to the logical error rate. 
We construct an error budget for $\Lambda$ (see Eqn.~(1) of the main text) by attempting to represent its inverse $\Lambda^{-1}$ as a linear function of the component errors, which we motivate by arguing that $\Lambda^{-1}$ is approximately linear in the component errors.  
For such a model, the fraction budgeted to each component is simply given by the weighted contribution of the component error, divided by quantity $\Lambda^{-1}$.  
However, $\Lambda^{-1}$ is not a perfectly linear function, and we discuss our approach to dealing with this.  
Our intent with the error budgeting is to determine what component error rates are necessary to implement a working demonstration of a surface code.  We can forecast how a small surface code might perform if run on a device with current error rates, and we can use the error budget to compare tradeoffs in component errors and make design decisions for future devices.

\subsection{A Description of a Component-Error Model for Simulations}
We simulate the repetition and surface code experiments in a simplified “circuit noise” model.  A circuit is constructed from component operations, including Clifford gates and related operations like initialization or measurement in the eigenbasis of a Pauli operator.  A circuit composed of these components can be simulated efficiently, and this set of instructions is sufficient to implement stabilizer codes such as repetition codes and surface codes.  

Noise in the circuit is simulated by sampling random Pauli errors and inserting them into the circuit according to the following probability model.  For each component, there is a “Pauli error channel,” which is a distribution over the possible Pauli errors to insert, including identity for no error (e.g. the distribution has 4 elements for single-qubit operation, or 16 for a two-qubit operation). For each component in the circuit, a Pauli error is sampled according to the distribution associated with that component, and this Pauli operator is inserted after the component.  Measurement errors are treated slightly differently, as follows.  The binary measurement result is flipped with a probability $p$, i.e. it goes through a classical binary symmetric channel instead of a Pauli channel.  For the circuits used in this work, when a qubit is measured, it is always reset before being used again; this means we do not assume that a measured qubit is left in the state consistent with a measurement result, because we unconditionally reset that qubit before using it again.

The effect of the randomly sampled Pauli errors that are injected into the simulated circuit is to change some of the measurement outcomes from their expected values.  For example, an $X$ (bitflip) error that occurs on a data qubit will be detected by the next syndrome circuits that interrogate this data qubit.  We collect the syndrome measurements and final data-qubit measurements in the simulation, and process them in the same way as the experiment using minimum-weight matching to infer a most likely location of errors.

Our simulations make some simplifying assumptions about the Pauli error channels.  First, we assume that each use of a component of the same type (e.g. every CZ gate) has the same error channel.  Of course, it would be straightforward to simulate different error channels for each gate in the circuit.  This would also be computationally efficient, but we opt to keep the number of parameters in the simulation relatively small.  Second, we further simplify error channels to be parameterized by a single scalar parameter.  The error channel for each gate or idle is a depolarizing channel parametrized by a single probability $p$ for any error to occur; for a single-qubit depolarizing channel, each of X, Y, or Z errors has probability $p/3$ to occur; for a two-qubit depolarizing channel, each of the 15 non-identity Paulis has probability $p/15$ to occur.  Each reset operation is followed by a quantum bitflip channel (random insertion of Pauli $X$), and each measurement operation is followed by a classical bitflip channel (random flip of the measurement bit).  All components (e.g. every CZ gate) have the same error channel, but different components can have different error probabilities (i.e. measurement error $p_{\mathrm{m}}$ can be distinct from the CZ error $p_{\mathrm{CZ}}$).

There are six types of component operations in our model, which are listed in Tab.\,\ref{table:errorrates}.  Since the error channel on each component has a single parameter, the noise in the simulator has six parameters.  We refer to these parameters collectively as a vector denoted $x$, which we use to relate the component-error probabilities to performance measures of the repetition and surface codes, such as logical-error probability or $\Lambda$, the ratio by which logical error improves when code distance is increased by 2.

\begin{table}
\caption{\label{table:errorrates}Error rates used in bit and phase flip simulations}
\begin{ruledtabular}
\begin{tabular}{c c c}
Component & Bitflip & Phaseflip \\
\hline
DD & 5.1e-2 & 4.1e-2 \\
CZ & 6.6e-3 & 6.6e-3 \\
M & 1.9e-2 & 1.9e-2 \\
R & 5.0e-3 & 5.0e-3 \\
H & 1.1e-3 & 1.1e-3 \\
I & 8.4e-4 & 5.8e-4 \\
\end{tabular}
\end{ruledtabular}
\end{table}

\subsection{Comparing Component-Error Simulations to the Experiments}

\begin{figure}
\includegraphics[width=85mm]{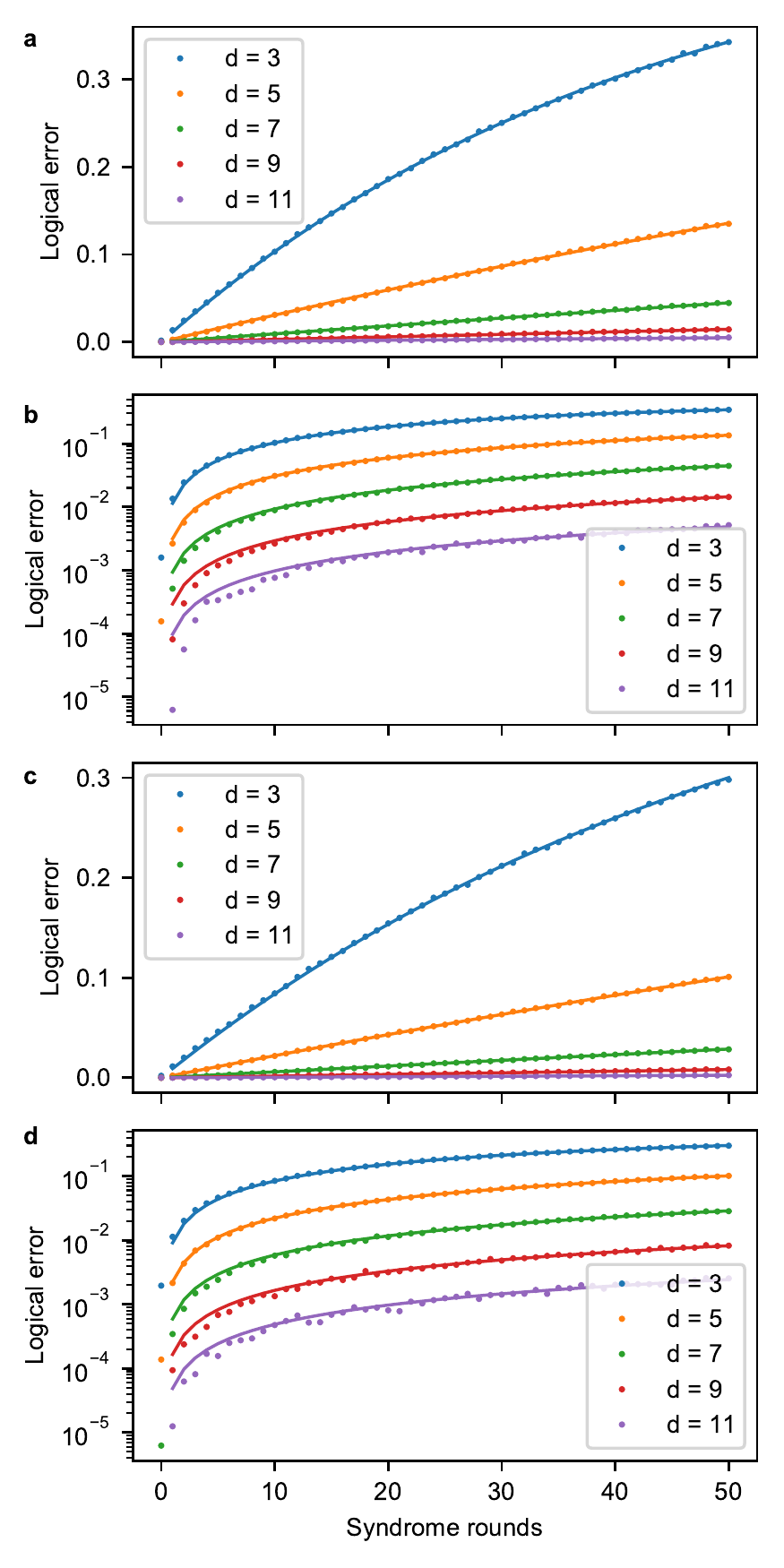}
\caption{\label{fig:sims1}  Simulations of logical-error probability for repetition codes using Pauli-channel noise calibrated to component errors measured in the device. \textbf{a,} Logical error vs. number of syndrome rounds for the bit flip code. \textbf{b,} Same data as panel \textbf{a} (bit flip code), plotted on a log-scaled vertical axis. \textbf{c,} Logical error vs. number of syndrome rounds for the phase flip code. \textbf{d,} Same data as panel \textbf{c} (phase flip code), plotted on a log-scaled vertical axis.}
\end{figure}

\begin{figure}
\includegraphics[width=85mm]{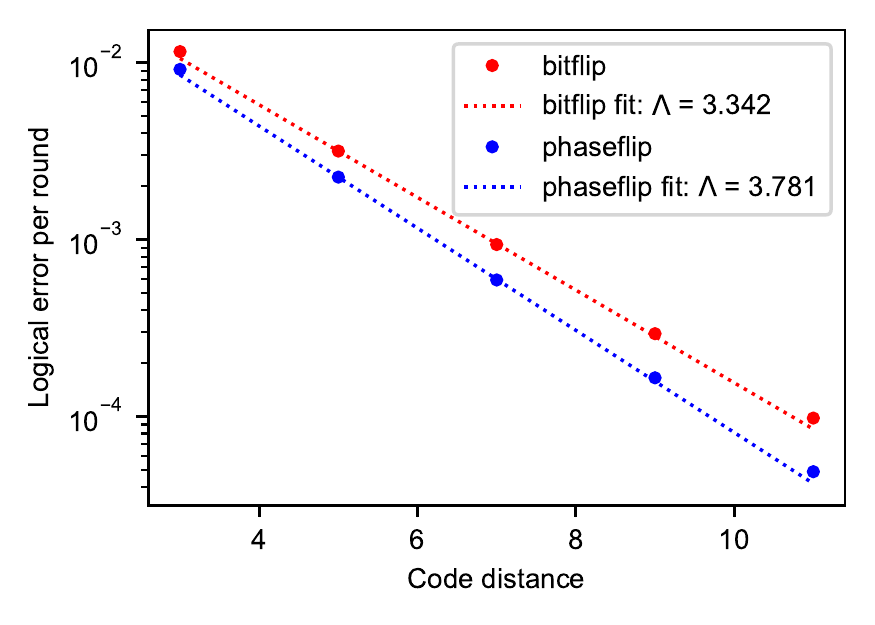}
\caption{\label{fig:sims2}  Logical error vs. code distance for the repetition codes, and a fit to estimate $\Lambda$ for the two codes.}
\end{figure}

To reproduce experimental conditions in the simplified simulator, we try to approximate the error rate in each component with data from benchmarking of those components.  The methods for characterizing error are:
\begin{itemize}
\item \textbf{Single- and two-qubit gates}: cross-entropy benchmarking~\cite{Bar2019}, averaged over the gates used in the experiment.  Averages treat one-qubit and two-qubit gates separately.

\item \textbf{Idle operations}: modeled as memoryless depolarizing channel with decay time constant given the by relevant experiment, meaning “$T_1$ decay” for the bitflip code and “$T_2$ decay” for the phaseflip code.  $T_1$ decay means initializing $\left\vert1\right\rangle$ and measuring probability of the state being $\left\vert1\right\rangle$ as a function of time; $T_2$ decay meanings initializing $\left\vert+\right\rangle$ and measuring decay of this state to the mixed state with time, while doing CPMG echoing to remove low-frequency phase noise (this dynamical decoupling is also done during idle operations in the phaseflip experiments).

\item \textbf{Reset and measurement}: These errors are difficult to distinguish; measurement error presents a noise floor for reset characterization. However, for simulation purposes, only the sum of the two error probabilities is important. We characterize reset by performing the reset gate between measurement pulses, preparing the qubit in $\left\vert0\right\rangle$ or $\left\vert1\right\rangle$; the error is the probability of finding $\left\vert1\right\rangle$ after reset. For measurement, we benchmark individual qubits by preparing $\left\vert0\right\rangle$ or $\left\vert1\right\rangle$ and immediately measuring, identifying the error probability. We also benchmark simultaneous readout on all the measure qubits and all the qubits, as in Ref.~\cite{arute2019quantum}.
\end{itemize}

It is important to note that the model is limited to only simulating Markovian Pauli channels.  The associated probability distributions are independent and identically distributed for each type of component.  Other important physical effects that we suspect to be present are not included in the model, such as leakage, cross-talk during gates, cosmic rays, parameter drift with time, or any other non-Markovian noise source.  The reason for choosing such a limited noise model is that it scales to large problem sizes and allows us to make forecasts of surface codes.  In future work, we will improve the simulations to incorporate approximations to effects like leakage that are still computationally efficient at large numbers of qubits.

The simulation conditions mirror the experiments in simulating bitflip and phaseflip error-correcting codes with the following parameters.  The values of component-error probabilities are those given in the main text, Fig. 4a.  The syndrome circuits are executed $n_{\mathrm{rounds}}$ times, for $n_\text{rounds}$ being every integer in the range [1,50].  At each value of $n_{\mathrm{rounds}}$, the simulation is executed $M$ = 160,000 times.  A logical error has occurred if the logical measurement at the end of an error-correction circuit gives an encoded qubit state different from the initial encoded state.  We count the number of simulated logical errors $m_e(n_{\mathrm{rounds}})$ at each value of $n_{\mathrm{rounds}}$, and the logical error probability is calculated as
\begin{equation}
P_{\mathrm{error}}(n_{\mathrm{rounds}}) = m_e(n_{\mathrm{rounds}}) / M.
\end{equation}
For each value of code distance $d \in \{3, 5, 7, 9, 11\}$, we determine the logical error rate $\epsilon_\text{logical}$ by fitting
\begin{equation}
P_{\mathrm{error}}(n_{\mathrm{rounds}}) ~= 0.5 \left[ 1 - \left(1 - 2 \epsilon_{\mathrm{logical}}\right)^{n_{\mathrm{rounds}}} \right]
\end{equation}
to the sampled data.  This fitting ansatz has the properties that $P_{\mathrm{error}}(n_{\mathrm{rounds}}=0) = 0$, it saturates as $P_{\mathrm{error}}(n_{\mathrm{rounds}} \rightarrow \infty) = 0.5$, and the error after one round $P_{\mathrm{error}}(1) = \epsilon_{\mathrm{logical}}$.  As in the main text, we calculate $\Lambda$ as the ratio by which logical error improves when increasing the code distance by 2:
\begin{equation}
\Lambda(d) = \epsilon_{\mathrm{logical}}(d) / \epsilon_{\mathrm{logical}}(d+2).
\end{equation}
The simulated logical error vs. number of syndrome rounds, and fits to this data, are shown in Fig.\,\ref{fig:sims1}. The simulated logical error rates match well but not perfectly to the experimental results.  Figure \ref{fig:sims2} shows the fitted logical error per round vs. code distance and fits to determine $\Lambda$.  The error rates are lower, and $\Lambda$ values are higher, than what is seen in the experiments.  We attribute this discrepancy to one of the assumptions of the simulator not holding in experiment.  For example, Section \ref{section:pij} discusses evidence for cross-talk errors happening during the experiment as well as long time correlations in detection events due to presence of leakage states in the data qubits.  Another possibility is that parameter drift during the experiment leads to higher error rates when running error correction than during the component benchmarking that determines the component error probabilities used in the simulation.  Said another way, this method of forecasting $\Lambda$ accounts for about 85\% of the error, because it predicts $\Lambda^{-1}$ values that are about 0.85 of the experimentally measured values, leaving weighted error contributions of about 15\% of the total not accounted for.  This method was also used to simulate the d=2 surface code, producing the “model” traces in Fig. 4c-d of the main text.

\subsection{Error Budgeting: Constructing a Linear Model Relating Component Errors to Inverse of Lambda}

The quantity $\Lambda$ is used to forecast logical error rate for a quantum code of a given size, so we extend this reasoning to determine what component error rates are needed to realize a target $\Lambda$ value.  We use the convention that $\Lambda$ is the factor by which logical error is suppressed by increasing code size, where $\Lambda > 1$ means logical error decreases when code size increases.  As a ratio, its inverse $\Lambda^{-1}$ has the same meaning (the factor by which logical error changes when code size increases one step).  Moreover, we argue that $\Lambda^{-1}$ is approximately a linear function of component errors.  As in the main text, we say that logical error rate is related to code distance $d$ by $\epsilon_{\mathrm{logical}}(d) \propto \Lambda^{-[(d+1)/2]}$ for $d$ odd.  It has been seen in numerical simulations with Pauli-channel noise~\cite{fowler2012towards, BSV2014} that for a single physical-error parameter $p$, $\epsilon_{\mathrm{logical}} \propto (p/p_{\mathrm{th}})^{[(d+1)/2]}$, where $p_{\mathrm{th}}$ is the threshold error rate for the chosen code and error model parameterized by $p$.  Hence, a naive comparison of the two approximate expressions would have $\Lambda^{-1} = p/ p_{\mathrm{th}}$, meaning that $\Lambda^{-1}$ is (approximately) linear in $p$.

For notational simplicity, denote the vector of component error rates as $x$ and let there be a function of component error rates $f(x)$ such that $\Lambda^{-1} = f$.  We will assume throughout that $f(0) = 0$, meaning $\Lambda$ approaches $\infty$ in the limit errors go to zero.  If $f(x)$ were a truly linear function in its arguments, we could calculate the gradient $g = \nabla f$ anywhere to determine $f$ exactly.  However, numerical simulations show that this is not the case, and the gradient changes for different choices of the point to linearize around.  Since we desire a linear model to form an error budget, we need to make a choice of how to do so; since $f(x)$ is not linear, there is no single “correct” answer.

Our approach is to treat $f(x)$ as if it was a second-order function in its arguments,
\begin{equation}
f(x) \approx g x + 0.5 x^{\intercal} H x,
\end{equation}
where $g$ is the gradient of $f$, $(H)_{ij} = \partial^2 f / \partial x_i \partial x_j$ is the Hessian matrix of $f$, and both are evaluated at $x \rightarrow 0^{+}$.  By doing so, we are saying that the second-order terms would capture enough of the nonlinearity in $f$ to provide a good approximation in the domain of interest.  We then exploit the following property.  For any second-order function $f$ with $f(0)=0$, there is a linear function given by the first-order Taylor series evaluated at a point $a/2$ such that this linear function coincides with the second-order function at $a$:
\begin{equation}
 \left( \nabla f \vert_{x = a/2} \right) a = g a + 0.5 a^{\intercal} H a = f(a)
\end{equation}
To make an error budget for the experimental component-error vector $x$ (values in Fig. 4a of the main text), we use simulations to numerically evaluate the gradient of $f$ at $x/2$, which determines the weights on the error components.  From the weights in this linear model, we can produce an estimate of $f = \Lambda^{-1}$ that shows the weighted contribution of each component error.  These results are summarized in Tab.\,\ref{table:bitflipbudget} and Tab.\,\ref{table:phaseflipbudget}.

\begin{table*}[t]
\caption{\label{table:bitflipbudget}\textbf{Error Budget for bit flip code.}}
\begin{ruledtabular}
\begin{tabular}{c c c c c }
Component & Error rate & Model weight & Contribution to $\Lambda^{-1}$ & Error-budget percentage \\
\hline
DD & 5.1e-2 & 3.5 & 0.179 & 58\% \\
CZ & 6.6e-3 & 11.7 & 0.077 & 25\% \\
M & 1.9e-2 & 1.6 & 0.030 & 10\% \\
R & 5.0e-3 & 1.6 & 0.008 & 3\% \\
H & 1.1e-3 & 3.4 & 0.004 & 1\% \\
I & 8.4e-4 & 6.6 & 0.006 & 2\% \\
\end{tabular}
\end{ruledtabular}
\end{table*}

\begin{table*}[t]
\begin{ruledtabular}
\caption{\label{table:phaseflipbudget}\textbf{Error budget for phase flip code.} *Note that “I” gates are assigned zero weight.  The term in the gradient of $\Lambda^{-1}$ for this component is actually a small negative number that depends on code distance, for example about -1 for $\Lambda$ between d=3 and d=5.  The reason this is negative is that “I” gates only appear on data qubits at the endpoints of the linear chain, and not across the data qubits like the other components.  This is why the derivative of $\Lambda^{-1}$ with respect to “I”-gate probability is negative: errors in this component affect d=3 more than d=5, and the trend continues to higher distances.  For the experimentally measured error rate in this component, it has negligible contribution to logical error and hence $\Lambda^{-1}$, so we choose to set its weight to zero for the purposes of an error budget.}
\begin{tabular}{c c c c c }
Component & Error rate & Model weight & Contribution to $\Lambda^{-1}$ & Error-budget percentage \\
\hline
DD & 4.1e-2 & 3.5 & 0.144 & 54\% \\
CZ & 6.6e-3 & 11.9 & 0.079 & 29\% \\
M & 1.9e-2 & 1.5 & 0.029 & 11\% \\
R & 5.0e-3 & 1.5 & 0.008 & 3\% \\
H & 1.1e-3 & 8.0 & 0.009 & 3\% \\
I & 5.8e-4 & 0* & 0 & 0\% \\
\end{tabular}
\end{ruledtabular}
\end{table*}

We see in these tables that the major source of logical error (more than 50\% of the budget) is idling error during the measurement and reset process.  This is simply due to $T_1$ decay times around 15\,$\mu$s and idle times (880\,ns during measurement and reset), leading to an error probability of 4--5\% during each such operation.  CZ gates and the combined effect of reset and measurement account for most of the remaining errors, with very small contributions from one-qubit gates and idle operations during gates.

\FloatBarrier  
\section{Probability \boldmath$p_{ij}$ of error-paired detection events}
\label{section:pij}

In this section, we discuss a technique that allows us to characterize error processes in repetition code experiments using correlations between detection events. We refer to this technique as the $p_{ij}$ {\it correlation matrix method}. We use it to estimate the probability $p_{ij}$ of conventional (e.g., bit or phase flips) and unconventional (e.g., leakage and crosstalk) error processes that produce pairs of detection events at the error graph nodes $i$ and $j$. 
We use this technique to produce in-situ diagnostics for QEC operation, and because it extracts detailed error information, it can also inform weights to the decoder.

\subsection{Error graph and correlation matrix \boldmath$p_{ij}$}

\begin{figure}
\centering
\includegraphics[width=\linewidth, trim =1cm 1.5cm 1cm 1.5cm,clip=true]{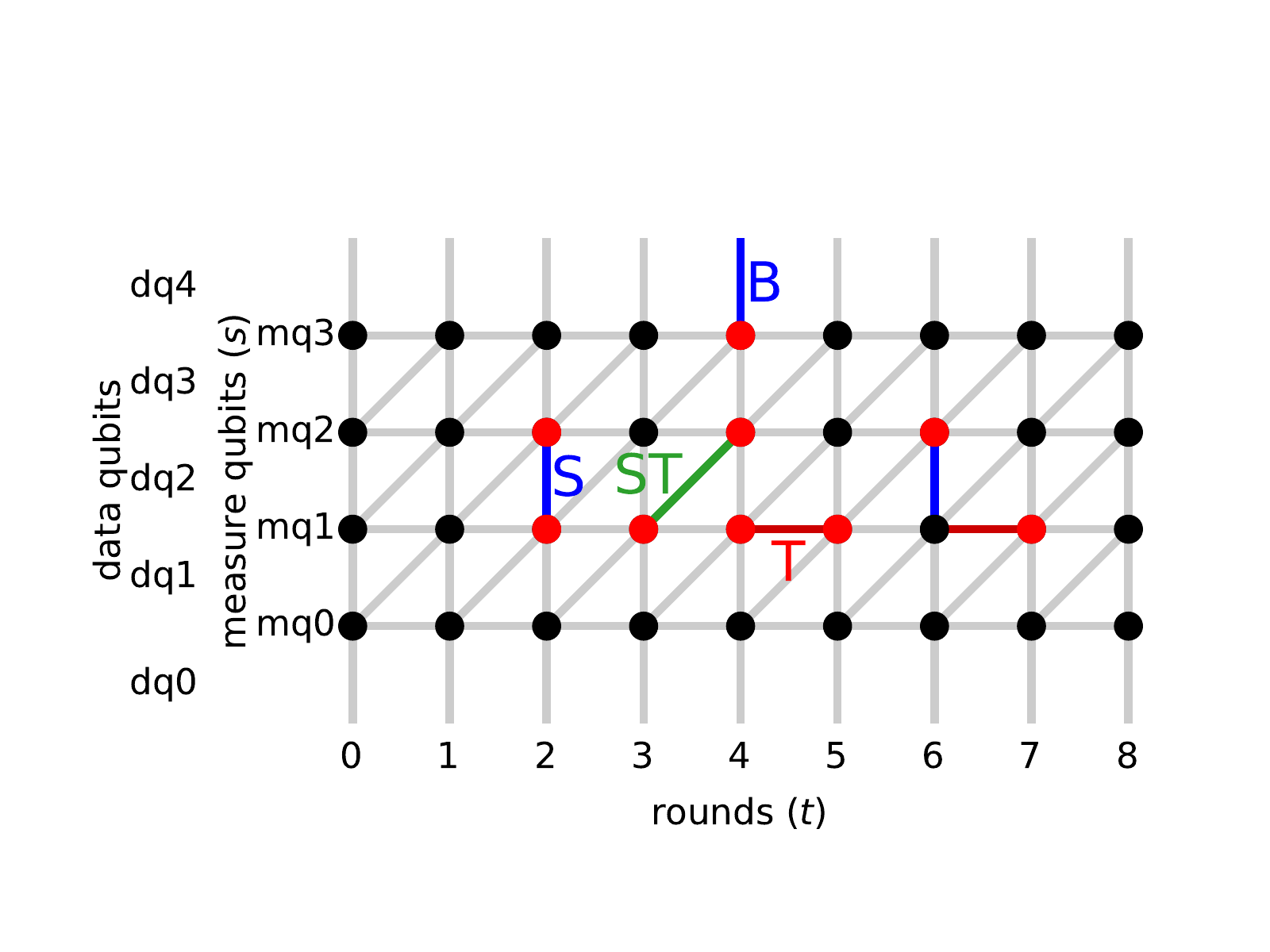}
\caption{{\bf Error graph and main edges.} An example of the error graph for $N_{\rm mq}=4$ measure qubits (5 data qubits) and $N_{\rm r}=8$ time rounds. The horizontal axis shows numbering of rounds ($t$ coordinate), the vertical axis shows numbering of measure qubits mq0--mq3 ($s$ coordinate). The dots denote the graph nodes; red dots indicate detection events. The vertical, horizontal and diagonal edges are denoted as Spacelike (S) [including the Boundary (B)],  Timelike (T) and Spacetimelike (ST) edges. Positions of data qubits dq0--dq4 (not used in the error graph) are indicated at the left.}
\label{fig:suppl-pij-error-graph}
\end{figure}

Figure~\ref{fig:suppl-pij-error-graph} shows an example of the error graph of a quantum bit-flip or phase-flip repetition code. It contains $(N_{\rm r}+1)N_{\rm mq}$ nodes (vertices), where $N_{\rm r}$ is the number of rounds ($0, 1, ... N_{\rm r}-1$) and $N_{\rm mq}$ is the number of measure qubits (the number of data qubits is then $d=N_{\rm mq}+1$, which is also the code distance). Each node $i$ corresponds to readout of a measure qubit (except for the last column of nodes -- see below) and can be associated with a pair of error graph coordinates: $i=\{s, t \}$, where $s=0, 1, ... N_{\rm mq}-1$ is the space-coordinate (measure qubit index) and $t=0, 1, ...N_{\rm r}$ is the time-coordinate (round number). The nodes can also be counted, e.g., in  the ``time-first'' manner,
    \begin{align}
i = t + (N_{\rm r}+1) s,
    \label{time-first}\end{align} 
or in the ``space-first'' manner, 
    \begin{align}
i = s + N_{\rm mq}\, t. 
    \label{space-first}\end{align} 

In each experiment, some of the nodes experience error detection events \cite{kelly2015state} (or simply ``detection events'') denoted by red dots in Fig.\,\ref{fig:suppl-pij-error-graph} (black dots denote absence of detection events). By definition, a detection event at node $i=\{ s, t \}$ occurs when the corresponding measurement result $m_{\{ s, t \}}$ is different from the previous measurement of the same qubit, $x_{\{ s, t \}} = m_{\{ s, t \}} \oplus m_{\{ s, t-1 \}}$, where $x_i=1$ means a detection event at node $i$, while $x_i=0$ means no detection event (here $\oplus$ denotes XOR). There are two exceptions to this rule. First, for the column with $t=0$, instead of non-existing $m_{\{s,-1\}}$ we use the parity of two neighboring data qubits in the initial state (if there is no error, we are supposed to get $x_{\{ s, 0 \}}=0$). The second special case is for the last column of nodes, $t=N_{\rm r}$, which does not correspond to a physical round (physical rounds are $t=0,1,...N_{\rm r}-1$); in this case, instead of non-existing $m_{\{s,N_{\rm r}\}}$, we use the parity of neighboring data qubit readouts at the end (after the round $N_{\rm r}-1$), so that $x_{\{s,N_{\rm r}\}}=0$ again indicates the expected no-error situation.

A decoder's task is to use detection events on the error graph to choose one of two given complementary initial states of data qubits (initial parities of neighboring data qubits are given, so the  decoder needs to determine only one bit of information). The decoder for this experiment used minimum-weight perfect matching algorithm \cite{kelly2015state, fowler2012towards, fowler2015minimum}, which connects detection events to each other (pairwise) or to a space-boundary.

In the conventional Pauli error model assumed by the decoder \cite{kelly2015state}, the detection events can be produced only in pairs, corresponding to the edges of the error graph (for the space-boundary edges, only one detection event near the boundary is produced). There are 3 types of such edges -- see Fig.\,\ref{fig:suppl-pij-error-graph}. Spacelike (S) edges connect nodes $\{s,t\}$ and $\{s+1,t\}$ (the boundary S-edges connect nodes $\{0,t\}$ and $\{N_{\rm mq}-1,t\}$ to the corresponding space-boundaries), timelike (T) edges connect nodes $\{s,t\}$ and $\{s,t+1\}$, and spacetimelike (ST, ``diagonal'') edges connect nodes  $\{s,t\}$ and $\{s+1,t+1\}$. In the conventional Pauli error model, a single physical error corresponds to an edge of the error graph.

Note that if two physical errors occur in edges sharing a node (see Fig.\,\ref{fig:suppl-pij-error-graph}), then there will be no detection event at this node: two detection events at the same node cancel each other. Therefore it is better to say that a physical error {\it flips color} (black$\leftrightarrow$red, $x_i\to 1-x_i$) of two nodes, instead of producing two detection events.

Now let us discuss how to find the probability $p_{ij}$ of a physical error, which flips colors of both nodes $i$ and $j$, using experimental statistics of detection events. From experimental data we see that such processes may occur not only when a pair of nodes is connected by a conventional edge on the error graph; therefore, we treat $i$ and $j$ as arbitrary nodes. However, we still assume that such pairs (edges) are uncorrelated with each other. In reality, sometimes there is a correlation between the edges (discussed later); so the assumption of the absence of correlation is a first approximation.

As mentioned above, $p_{ij}$ denotes the probability that two nodes $i$ and $j$ flip color simultaneously. These nodes can also flip color because of other edges connected to $i$ and $j$ separately. However, it is important that these additional flips are independent (uncorrelated) for $i$ and $j$ because they are caused by different physical errors. Therefore, we can consider three uncorrelated processes: node $i$ flips color ($x_i \to 1-x_i$) with some probability $p_i$, similarly node $j$ flips color with probability $p_j$, and both nodes flip  color with probability $p_{ij}$. Since we start with the black color ($x_i=x_j=0$), the joint probabilities $P(x_i, x_j)$ of detection or no detection events at nodes $i$ and $j$ are

    \begin{subequations}
\label{P-xi-xj}
\begin{align}
  P(0,0) = &\; (1-p_{ij})(1-p_i)(1-p_j) + p_{ij}p_ip_j , \\ 
    P(0,1) =&\; (1-p_{ij})(1-p_i)p_j + p_{ij}p_i(1-p_j) ,  \\
    P(1,0) =&\; (1-p_{ij})p_i(1-p_j) + p_{ij}(1-p_i)p_j , \\
    P(1,1) =&\; (1-p_{ij})p_ip_j + p_{ij}(1-p_i)(1-p_j) .
\end{align}
    \end{subequations} 
    
These formulas have obvious meaning, describing combinations of the three processes occurring or not occurring. Note that $P(0,0)+P(0,1)+P(1,0)+P(1,1)=1$. The relations (\ref{P-xi-xj}) can also be expressed via the fractions of the detection events (often abbreviated as DEF: detection event fraction) for each node, $\langle x_i\rangle =P(1,0)+P(1,1)$ and $\langle x_j\rangle =P(0,1)+P(1,1)$, and the probability of both detection events, $\langle x_i x_j \rangle =P(1,1)$, which gives

    \begin{subequations}
\label{DEFs-via-pij}
\begin{align}
& \langle x_i \rangle = p_i\left(1-p_{ij}\right) + \left(1-p_i\right)p_{ij}, \\
& \langle x_j \rangle = p_j\left(1-p_{ij}\right) + \left(1-p_j\right)p_{ij},\\
& \langle x_ix_j \rangle = p_{ij}\left(1-p_i\right)\left(1-p_j\right) + \left(1-p_{ij}\right)p_ip_j.
\end{align}
    \end{subequations}

Solving these equations for $p_{ij}$, $p_i$, and $p_j$,  we obtain

    \begin{align}
& p_{ij} = \frac{1}{2} - \frac{1}{2}\sqrt{1 - \frac{4 \left( \langle x_ix_j\rangle - \langle x_i \rangle\langle x_j \rangle \right) }{1 - 2\langle x_i\rangle - 2\langle x_j \rangle + 4\langle x_ix_j \rangle}} \, ,
    \label{pij-result}  \\
& p_i = \frac{\langle x_i \rangle - p_{ij}}{1-2p_{ij}}\, ,\;\;\; p_j = \frac{\langle x_j \rangle - p_{ij}}{1-2p_{ij}}\, .
    \label{pi-and-pj} \end{align}

 We can think about $p_{ij}$ as a symmetric matrix, $p_{ji}=p_{ij}$, with indices corresponding to the nodes ordered either in the ``time-first'' way (\ref{time-first}) or in the ``space-first'' way (\ref{space-first}) -- see Figs.\ \ref{fig:suppl-pij-time-first} and \ref{fig:suppl-pij-space-first} discussed later. Formally, in Eqn.\,(\ref{pij-result}) the diagonal elements are the detection fractions, $p_{ii}=\langle x_i\rangle$; however, we usually set them to zero, $p_{ii}\equiv 0$, for clarity of graphical presentation.

Note that in the experimentally relevant case when $p_{ij}\ll 1/4$, Eqn.\,(\ref{pij-result}) can be approximated as ($i\neq j$) 

    \begin{align}
     p_{ij} \approx \frac{ \langle x_i x_j\rangle - \langle x_i \rangle\langle x_j \rangle }{(1 - 2\langle x_i\rangle)(1 - 2 \langle x_j\rangle)}\, .
    \label{pij-approx}\end{align}
    
Equation (\ref{pij-approx}) for $p_{ij}$ is Eqn.\,(2) of the main text.
This form shows a clear relation of $p_{ij}$ to the covariance $\langle x_ix_j\rangle - \langle x_i \rangle\langle x_j \rangle$; however, the correction due to the denominator is typically quite significant. For example, for $\langle x_i \rangle \simeq \langle x_j \rangle \simeq 0.11$ (see Fig.\,1 of the main text), the denominator in Eqn.\,(\ref{pij-approx}) is about 0.6. The approximation (\ref{pij-approx}) slightly overestimates Eqn.\,(\ref{pij-result}), the correction factor is roughly $(1-3p_{ij})$. 

Equation (\ref{pij-result}) allows us to find accurate individual error probabilities for S, T, and ST edges of the error graph, which are needed for the minimum-weight decoder. However, there is an important exception: the error probability for a boundary S-edge cannot be obtained in this way because it contains only one node. 
To find the error probability $p_{i\,{\rm B}}$ for a boundary edge from node $i$, we use Eqn.\,(\ref{pi-and-pj}), $p_{i,\Sigma} = (\langle x_i \rangle - p_{i\,{\rm B}})/(1-2p_{i\,{\rm B}})$, in which the ``individual flip'' probability $p_{i,\Sigma}$ is calculated from already calculated error probabilities for S, T, and ST edges connected to the node $i$. We essentially sum up the known error probabilities of the connected edges and find  the missing error probability (due to the boundary edge) to bring the sum to the DEF $\langle x_i\rangle$. Note, however, that it is not a simple sum of the probabilities because of the ``color flipping'' procedure, so that the errors $p_{ij_1}$, $p_{ij_2}$, ... $p_{ij_k}$ due to $k$ connected edges produce the total flip probability

    \begin{align}
& p_{i,\Sigma} = g(p_{ij_k}, ... \, g(p_{ij_3},  g(p_{ij_2},p_{ij_1})) ...), 
    \label{p-i-sum}\\
& g(p, q)\equiv p(1-q)+(1-p)q = p+q-2pq.
    \end{align}
Thus, after finding $p_{i,\Sigma}$, we calculate the boundary S-edge probability as
\begin{align}
    p_{i\,{\rm B}} = \frac{\langle x_i\rangle -p_{i,\Sigma}}{1-2p_{i,\Sigma}}\, . 
\label{pij-boundary}\end{align}

Note that this procedure for boundary edges assumes that error processes corresponding to different edges are uncorrelated. In reality this is not a very good assumption (this is why we are actually using a slightly different procedure for boundary edges). A natural way to estimate the effect of correlation between the edges is to use Eqn.\,(\ref{p-i-sum}) for a node $i$ not close to a boundary, summing up the contributions from all connected edges and then comparing the result with the DEF $\langle x_i\rangle$. Doing this test for the phase-flip experiment, we typically find a relative inaccuracy of about 4\% (median value), which indicates a reasonably small but still nonzero correlation between the main edges (for the bit-flip experiment the median relative inaccuracy is about 9\%).
A natural way of thinking about positive correlations between the edges is to assume that some error processes flip color of 4, 6, ... nodes on the error graph, so that the same process increases $p_{ij}$ for several pairs of nodes (this also produces unconventional edges on the error graph reported by $p_{ij}$). 
To study correlations between edges, we have generalized the method of $p_{ij}$ to 3-point and 4-point correlators (essentially the ``hyperedges''), extending the approach of Eq.\,(\ref{P-xi-xj}) to account for more nodes and more error processes. This generalization will be described in a future publication.

\subsection{Fluctuations of the \boldmath$p_{ij}$ elements}

When evaluating Eqn.\,\eqref{pij-result} using experimental data, the $p_{ij}$ values exhibit statistical fluctuations because the averages $\langle x_ix_j\rangle$, $\langle x_i \rangle$,  and $\langle x_j \rangle$ are estimated from a large but finite number $N_{\rm expt}$ of experimental realizations (typical values of $N_{\rm expt}$ are between $10^3$ and $10^5$). In this section we estimate the standard deviation $\sigma_{p_{ij}}$ of statistical fluctuations of the $p_{ij}$ elements.

For the estimate, let us use the approximation (\ref{pij-approx}) and assume the usual experimental case when $\langle x_i\rangle \ll 1$, $\langle x_j\rangle \ll 1$, and $p_{ij} \ll 1$. Then the effect of the denominator fluctuations is negligible in comparison with fluctuations of the numerator (covariance $C_{ij}$), so 
    \begin{equation}
    \sigma_{p_{ij}} \approx \frac{ \sigma_{ C_{ij}} }{(1 - 2\langle x_i\rangle)(1 - 2 \langle x_j\rangle)}\, , \,\,\, C_{ij}= \langle x_i x_j\rangle - \langle x_i \rangle\langle x_j \rangle .
    \label{sigma-pij-1}\end{equation}
Using the form $C_{ij}=\langle(x_i-\langle x_i\rangle )(x_j-\langle x_j\rangle )\rangle$ and using in it true averages $\langle x_i \rangle$ and  $\langle x_j \rangle$ instead of averages over $N_{\rm expt}$ realizations (the effect of the change is negligible), we find
    \begin{equation}
    \sigma_{C_{ij}} = \sqrt{ {\rm Var}[(x_i-\langle x_i\rangle )(x_j-\langle x_j\rangle )]/N_{\rm expt}}\, .
    \end{equation} 
The variance here is $\langle (x_i-\langle x_i\rangle )^2(x_j-\langle x_j\rangle )^2 \rangle -C_{ij}^2$, in which the first term can be rewritten after some algebra as $C_{ij}(1-2\langle x_i\rangle)(1-2\langle x_j\rangle)+\langle x_i\rangle \langle x_j\rangle(1-\langle x_i\rangle)(1-\langle x_j\rangle)$, using the properties $x_i^2=x_i$ and $x_j^2=x_j$. Inserting this form into Eqn.\,(\ref{sigma-pij-1}) and using $C_{ij}/[(1 - 2\langle x_i\rangle)(1 - 2 \langle x_j\rangle)]\approx p_{ij}$, we obtain 

    \begin{equation}
    \sigma_{p_{ij}} \approx \frac{ \sqrt{ p_{ij}(1-p_{ij})+ \frac{\displaystyle \langle x_i\rangle \langle x_j\rangle (1-\langle x_i\rangle)(1-\langle x_j\rangle) }{\displaystyle (1 - 2\langle x_i\rangle)^2(1 - 2 \langle x_j\rangle)^2}}}{\sqrt{N_{\rm expt}} }\, .
    \label{sigma-pij-2}\end{equation}

Note that the first and second terms in the numerator of Eqn.\,(\ref{sigma-pij-2}) have a clear meaning and can be obtained separately. When $p_{ij}$ is well above the statistical noise floor, $\sigma_{p_{ij}}$ mainly comes from fluctuation of the number of realizations, in which the edge error (color flipping event) has occurred: $N_{\rm expt} p_{ij}\pm \sqrt{N_{\rm expt} p_{ij} (1-p_{ij})}$, as follows from the binomial statistics. It is easy to see that this leads to the first term in Eqn.\,(\ref{sigma-pij-2}). The second term is the noise floor, coming from the fluctuations of $\langle x_i\rangle$, $\langle x_j\rangle$, and $\langle x_i x_j \rangle$ when $p_{ij}=0$. It can be obtained, e.g., by considering the number of realizations with $x_i=1$: $N_{x_i=1}=N_{\rm expt} \langle x_i\rangle \pm \sqrt{N_{\rm expt} \langle x_i\rangle (1-\langle x_i\rangle)}$, number of realizations with $x_i=x_j=1$: $N_{x_i=x_j=1}=N_{x_i=1} \langle x_j\rangle \pm \sqrt{N_{x_i=1} \langle x_j\rangle (1-\langle x_j\rangle)}$ (with uncorrelated $\pm$), and realizations with $x_i=0$ and $x_j=1$: $N_{x_i=0,\, x_j=1}=(N_{\rm expt}-N_{x_i=1}) \langle x_j\rangle \pm \sqrt{(N_{\rm expt}-N_{x_i=1}) \langle x_j\rangle (1-\langle x_j\rangle)}$ (also with uncorrelated $\pm$). Then calculating the apparent value of the covariance $C_{ij}$ and using it in Eqn.\,(\ref{sigma-pij-1}), we obtain the noise floor, which gives the second term in Eqn.\,(\ref{sigma-pij-2}).

As a final simplification, let us neglect the factors $(1-p_{ij})$ and $(1-\langle x_i\rangle )(1-\langle x_j\rangle )$ in Eqn.\,(\ref{sigma-pij-2}) (this slightly increases $\sigma_{p_{ij}}$, so we are on the safe side), thus obtaining
    \begin{equation}
    \sigma_{p_{ij}} \approx \frac{1}{{\sqrt{N_{\rm expt}}} } \sqrt{ p_{ij}+ \frac{ \langle x_i\rangle \langle x_j\rangle  }{ (1 - 2\langle x_i\rangle)^2(1 - 2 \langle x_j\rangle)^2}}\, .
    \label{sigma-pij-3}\end{equation}

In our repetition phase-flip code experiments, we have {$N_{\rm expt}=76,000$} realizations and the detection error fractions are $\langle x_i \rangle \simeq \langle x_i \rangle \simeq 0.11$ (slightly bigger, $\simeq 0.12$ in the bit-flip experiments). Thus, the  standard deviation of the experimental $p_{ij}$ values that are nominally zero (noise floor) is roughly
\begin{align}
\sigma_{p_{ij}}\simeq 6\times 10^{-4}.
\label{sigma-pij-4}\end{align}

In particular, this is the noise floor seen in the $p_{ij}$ matrix plots shown in  Figs.~\ref{fig:suppl-pij-time-first} and~\ref{fig:suppl-pij-space-first}. Additional averaging over the rounds leads to even smaller noise floor ($<2\times 10^{-4}$) in Fig.\,2(c) of the main text.

\subsection{Experimental results for \boldmath$p_{ij}$} 

\begin{figure*}[h]
\centering
\includegraphics[width=0.9\paperwidth, trim =1cm 1cm 1cm 1cm,clip=true]{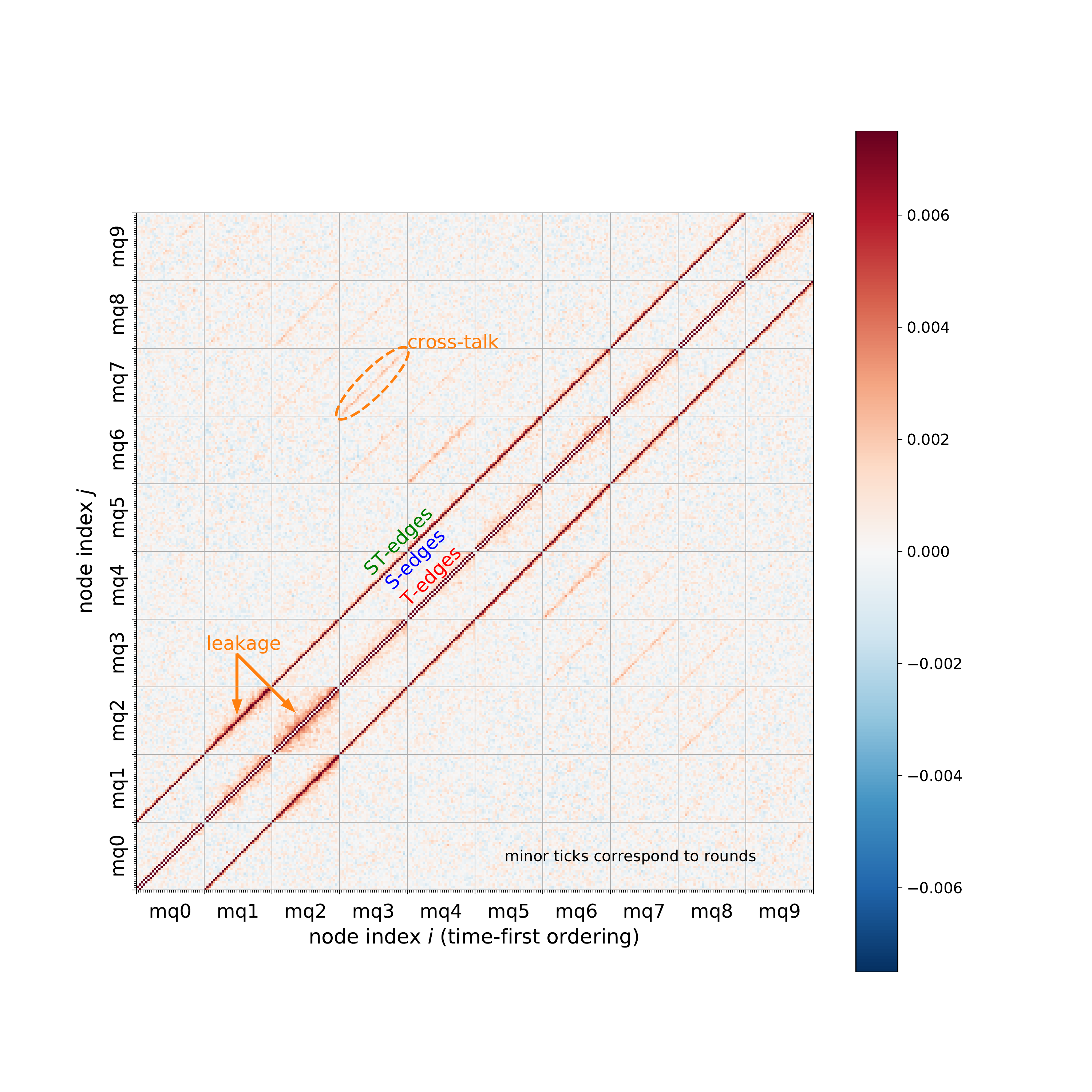}
\caption{{\bf Correlation matrix \boldmath$p_{ij}$.} A graphical representation of the 310$\times$310 symmetric matrix $p_{ij}$ [Eqn.\,(\ref{pij-result})] for a phase-flip repetition code experiment with $N_{\rm mq} =10$ measure qubits (11 data qubits) and $N_{\rm r} = 30$ rounds. The color of each pixel depicts the probability $p_{ij}$ for an error process involving error graph nodes $i$ and $j$. The nodes are ordered in the ``time-first'' fashion, Eqn.\,\eqref{time-first}, with 10 blocks (separated by grid lines) corresponding to measure qubits (mq0, mq1, ... mq9) and 31 ticks within each block corresponding to time rounds (from $t=0$ to $t=N_{\rm r}$). The main features are the diagonal lines corresponding to T, S, and ST edges, which are shifted from the main diagonal by 1, 31, and 32 pixels, respectively (ST line is more faint than T and S lines). Additional features are reddish (``dirty'') patches near S and T lines, which are due to leakage to state $|2\rangle$ in data qubits, and also short parallel lines (``scars'') due to crosstalk. Note that the color bar ranges to 0.007, while probabilities for S and T edges are above this truncation.}
\label{fig:suppl-pij-time-first}
\end{figure*}

\begin{figure*}[h]
\centering
\includegraphics[width=0.9\paperwidth, trim =1cm 1cm 1cm 1cm,clip=true]{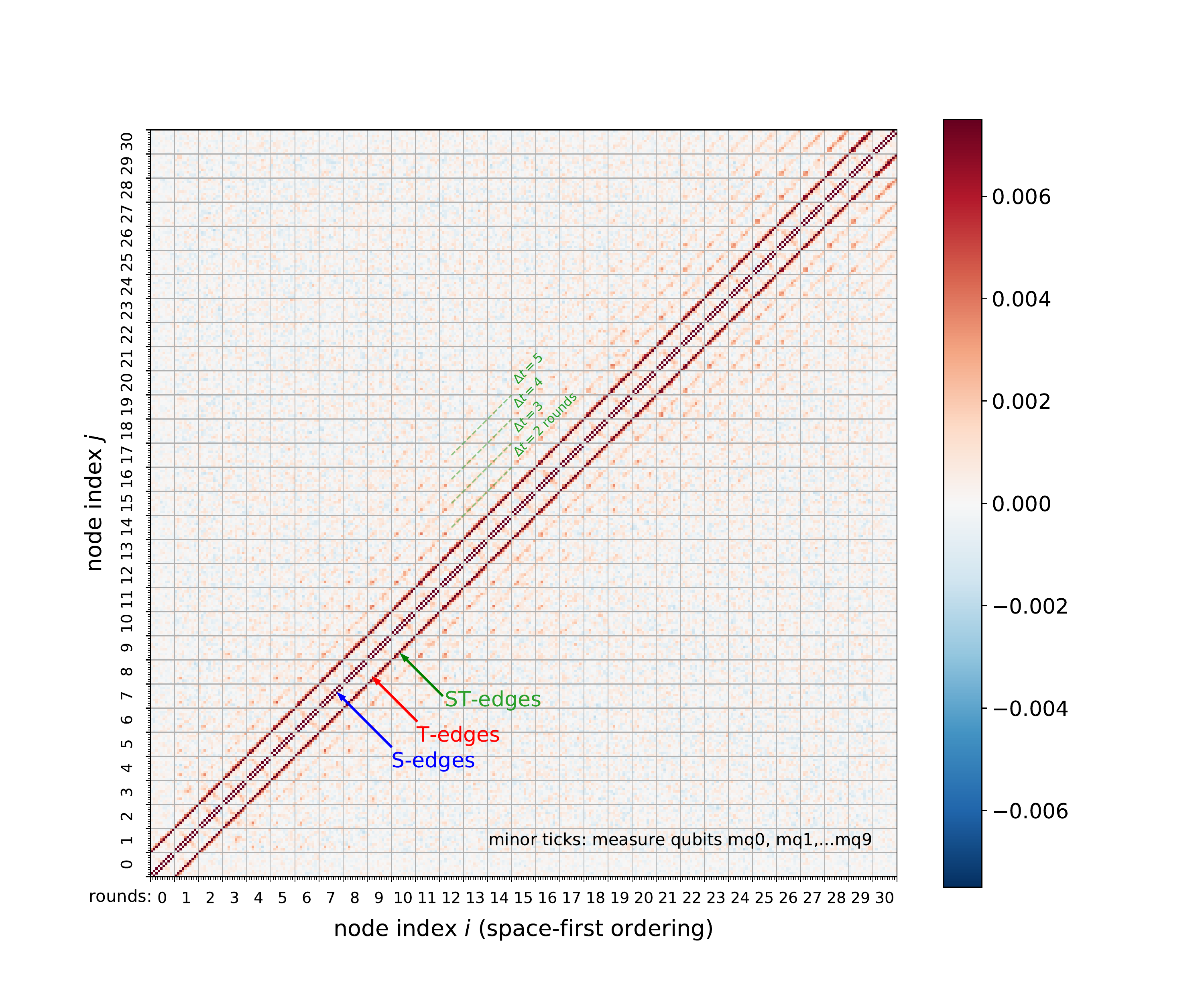}
\caption{{\bf Matrix \boldmath$p_{ij}$ in space-first node ordering.} The figure shows the same data as in Fig.\,\ref{fig:suppl-pij-time-first}, but with the nodes ordered in the ``space-first'' fashion of Eqn.\,(\ref{space-first}). Each axis contains $N_{\rm r}+1=31$ blocks with $N_{\rm mq}=10$ points (ticks) within each block. The lines for S, T, and ST edges are shifted from the main diagonal by 1, 31, and 32 pixels, respectively. Short dashed lines correspond to 2T, 3T, ... edges, which connect nodes separated by  $\Delta t=2$, 3, ...  rounds. The well-visible diagonal stripes indicate the presence of long-time correlations in detection events lasting for over 5 rounds.
}
\label{fig:suppl-pij-space-first} 
\end{figure*}

Figure \ref{fig:suppl-pij-time-first} shows the correlation matrix $p_{ij}$ for a phase-flip code experiment with 21 qubits ($N_{\rm mq} = 10$ measure qubits and 11 data qubits) and $N_{\rm r} = 30$ rounds. In this particular experiment, no cosmic rays events were detected, so no data was discarded from $N_{\rm expt}=76,000$ runs. The error graph nodes $i$ and $j$ are ordered in the ``time-first'' way given by Eqn.\,(\ref{time-first}). Figure \ref{fig:suppl-pij-time-first}  contains 310$\times$310 pixels, with the color of each pixel determined by the value of the corresponding $p_{ij}$ element. Each axis contains $N_{\rm mq}=10$ blocks (see grid lines) corresponding to 10 measure qubits indicated on the axes; each block contains $N_{\rm r}+1=31$ points (small ticks on the axes) corresponding to time rounds.

We see that most pixels in Fig.\,\ref{fig:suppl-pij-time-first} (which are away from the features discussed below) have values close to zero. The fluctuations are consistent with the expected noise floor given by Eqn.\,(\ref{sigma-pij-4}). The figure is symmetric across the main diagonal (which runs bottom-left to top-right) because $p_{ji}=p_{ij}$. The values on the main diagonal are set to zero.

\begin{figure*}[h]
\centering
\includegraphics[width=0.9\paperwidth, trim =3cm 0.1cm 0cm 0cm,clip=true]{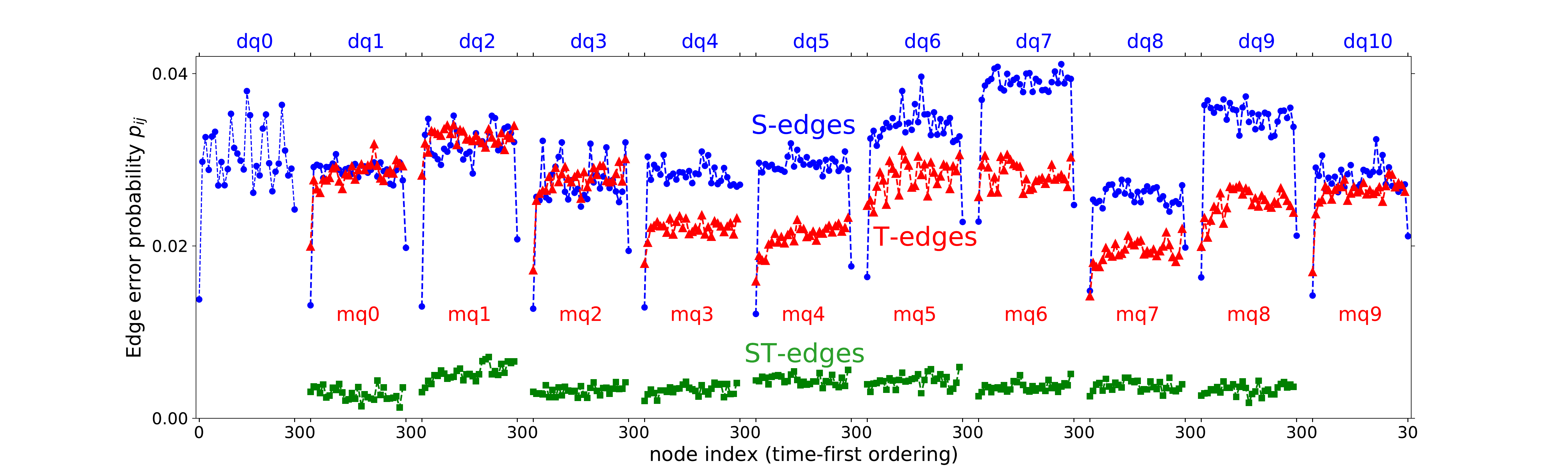}
\caption{{\bf S, T, and ST errors.} The plot shows the error probabilities $p_{ij}$ for S (spacelike), T (timelike), and ST (spacetimelike) edges for the data in Fig.\,\ref{fig:suppl-pij-time-first} (phase-flip code, 10+11 qubits, 30 rounds). For S-edges (blue symbols) the corresponding data qubits dq0--dq10 are indicated at the top, 31 points within each block correspond to rounds. The S-edge probabilities for boundary data qubits dq0 and dq10 are calculated using Eqs.\ (\ref{p-i-sum})--(\ref{pij-boundary}). For T-edges (red symbols), the corresponding measure qubits mq0--mq9 are indicated below the red symbols, each block contains 30 points. ST-edges (green symbols) are positioned in the same way as S-edges (without boundaries), with 30 points per block. Lines are  a guide for the eye.}
\label{fig:suppl-pij-STST}
\end{figure*}

The most visible features are 4 diagonal lines (2 from each side of the main diagonal), which correspond to S and T edges of the error graph: the T-edge line contains pixels next to the main diagonal, while S-edge line is $N_{\rm r}+1$ pixels away from the main diagonal. The color scale for S and T lines is saturated because the values of $p_{ij}$ for these lines are around 0.03; they are shown in Fig.\,\ref{fig:suppl-pij-STST} discussed in more detail below.
There is also a less visible line in Fig.\,\ref{fig:suppl-pij-time-first} next to the S-line (one pixel farther, $N_{\rm r}+2$, from the main diagonal), which corresponds to ST edges. The typical values of $p_{ij}$ for the ST-line are around 0.004. Another well-visible feature in Fig.\,\ref{fig:suppl-pij-time-first} is a reddish ``dirt'' near S and T lines for qubits mq1 and mq2 and to a less extent for some other qubits; we attribute this feature to leakage to state $|2\rangle$ in a data qubit. One more feature is short lines (``scars'') parallel to the main diagonal, which we attribute to  crosstalk. The leakage and crosstalk are discussed later.

{\bf S, T, and ST edges}. In the conventional theory of the repetition QEC code, the errors are associated only with S, T, and ST edges. The elements of $p_{ij}$ show the probabilities of these errors individually for each edge on the error graph. We emphasize that these probabilities are obtained {\it in situ}, during the actual operation of the code, in contrast to estimates based on qubit coherence and gate fidelities. 

\begin{figure*}[h]
\centering
\includegraphics[width=0.8\paperwidth, trim =0cm 0cm 0cm 0cm,clip=true]{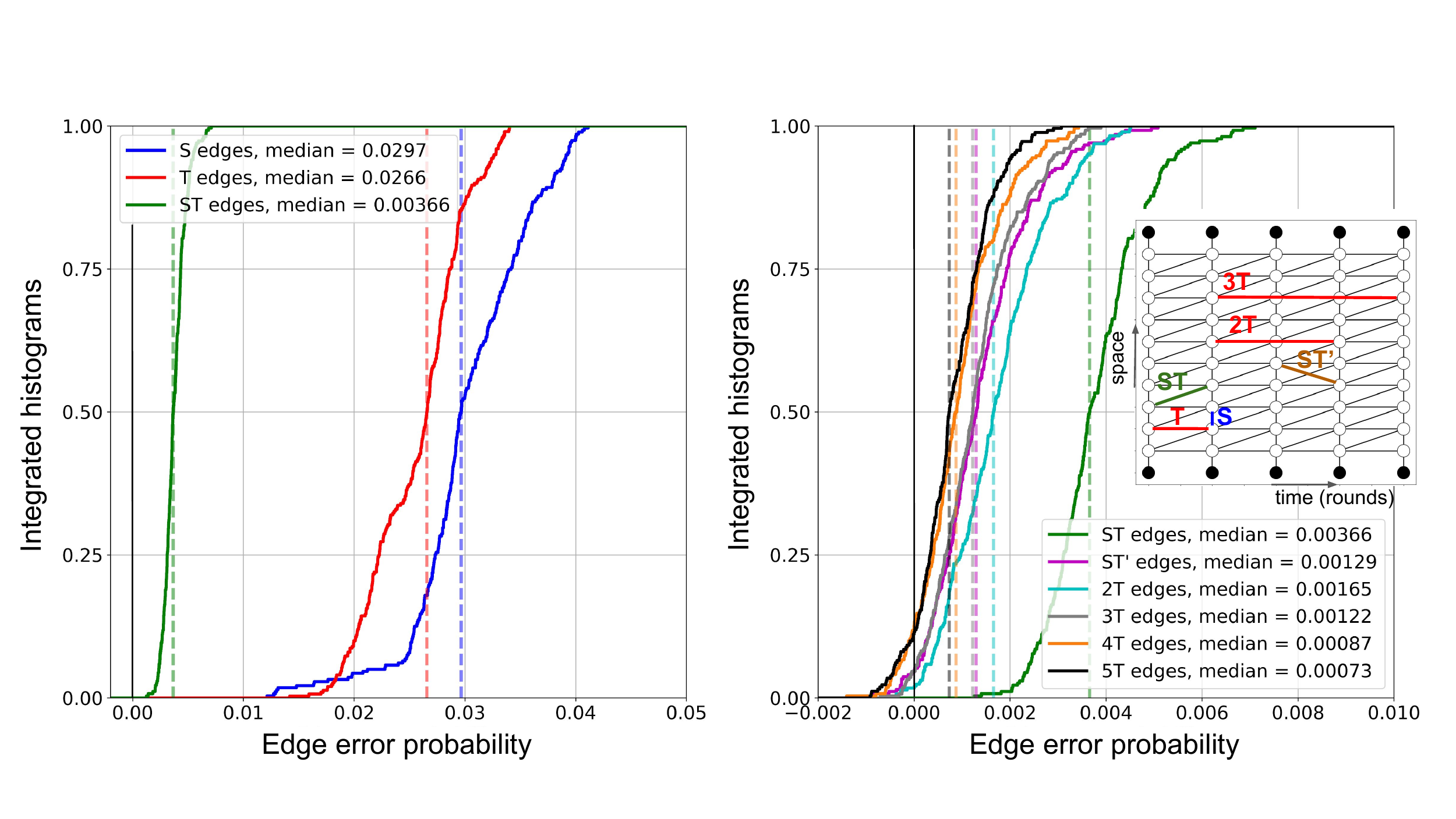}
\caption{{\bf Integrated histograms of edge error probabilities.} Left panel: Integrated histograms for the error probabilities $p_{ij}$ of the conventional edges: S (blue line), T (red line), and ST (green line). Median values are indicated by vertical dashed lines and shown in the legend. Right panel: Integrated histograms for ST edges and several unconventional edge types: other-direction spacetimelike edges ST$'$ and long timelike edges 2T, 3T, 4T, and 5T spanning 2, 3, 4, and 5 rounds. The inset illustrates the edge types on the error graph. Median values (dashed lines) are listed in the legend. 
}
\label{fig:suppl-pij-histogram}
\end{figure*}

As expected from the conventional theory, S, T, and ST edges are the main features in Fig.\,\ref{fig:suppl-pij-time-first}. The values of $p_{ij}$ elements for these edges are shown in Fig.\,\ref{fig:suppl-pij-STST} by blue markers for S-edges, red markers for T-edges, and green markers for ST-edges; the lines are a guide for the eye. The S-edge error probabilities for the boundary edges (denoted dq0 and dq10 in Fig.\,\ref{fig:suppl-pij-STST}) are calculated using Eqs.\ (\ref{p-i-sum})--(\ref{pij-boundary}); we see that their values are consistent with other S-edges. Each block of blue markers corresponds to a particular data qubit (indicated at the top), markers within a block correspond to time rounds (from 0 to 30, see the horizontal axis). Note that S-edge probabilities for rounds $t=0$ and $t=30$ are significantly smaller than for other rounds (emphasizing the need of many rounds in an experiment). This is because S-edge errors in our phase-flip code are mainly due to dephasing of data qubits during readout and reset (or due to energy relaxation for a bit-flip code), while the special rounds $t=0$ and $t=N_{\rm r}$ do not have these parts of the cycle. For other rounds, the error probability $p_{ij}$ can be crudely estimated as $\tau/2T_2$, where $\tau$ is the readout-and-reset time (expected contribution from CZ gates is significantly smaller). In our experiment, $\tau=0.88\, \mu$s and on average $T_2\simeq 16\, \mu$s, which gives $\tau/2T_2\simeq 0.028$. We see that $p_{ij}$ values for S edges (blue symbols) are close to this estimate, though they are different for different data qubits, mostly reflecting variation in $T_2$ times and also having contributions from gate errors. The integrated histogram for the S-edges is shown by the blue line in the left panel of Fig.\,\ref{fig:suppl-pij-histogram}; the median $p_{ij}$ value is 
    \begin{align}
    p_{ij}^{\rm S-edge,\, median} \approx 3.0 \times 10^{-2}.
    \end{align}

The T-edge errors (red symbols in Fig.\,\ref{fig:suppl-pij-STST}) are grouped in blocks corresponding to measure qubits indicated below the red symbols. The T-edge errors are expected to come mainly from the readout errors, but there are also contributions from the gate errors and reset error. Our median readout error is around 0.018; however, the $p_{ij}$ values are considerably higher, with the median value (see the integrated histogram in Fig.\,\ref{fig:suppl-pij-histogram}) of
   \begin{align}
    p_{ij}^{\rm T-edge,\, median} \approx 2.7\times 10^{-2}. 
    \end{align}
The error probabilities for ST edges (green symbols in Fig.\,\ref{fig:suppl-pij-STST}) are much lower than for S or T edges; they are supposed to come mainly from CZ gate errors. The integrated histogram in Fig.\,\ref{fig:suppl-pij-histogram} (green line) shows for ST edges the median value of
   \begin{align}
    p_{ij}^{\rm ST-edge,\, median} \approx 3.7 \times 10^{-3}.
    \end{align}

{\bf Unconventional edges}. Figure \ref{fig:suppl-pij-time-first} clearly shows that in contrast to what is expected from the conventional QEC theory, some correlations between the detection events correspond to error graph edges different from the S, T, and ST types. In particular, there are significantly non-zero $p_{ij}$ values near the lines corresponding to T and S edges, separated from them by a few rounds. The integrated histogram for some types of these edges is shown in the right panel of Fig.\,\ref{fig:suppl-pij-histogram}). As illustrated by the inset, with ST$'$ we denote the ``diagonal'' edges similar to the ST edges, but going into the other direction. With 2T, 3T, etc.\ we denote the edges spanning 2, 3, etc.\ rounds for the same measure qubit. We see that out of the unconventional edges, 2T edges have the highest typical probability (the median of $1.7\times 10^{-3}$), which is still more than twice smaller than the typical ST-edge probability. A relatively small probability of unconventional edges indicates a high quality of the experiment. Note that before the qubit reset \cite{mcewen2020} was implemented, the unconventional-edge probabilities were much higher, with 2T probabilities exceeding ST probabilities.

The negative values of $p_{ij}$ for a small fraction of unconventional edges shown in Fig.\,\ref{fig:suppl-pij-histogram} are consistent with the statistical noise level (\ref{sigma-pij-4}). Note, however, that in some cases, for example, for 2T edges in a high-quality bit-flip experiment, the $p_{ij}$ values can actually be slightly negative. This can be understood using Eqn.\,(\ref{pij-approx}) as a negative correlation. Indeed, a negative correlation between the nodes can be caused by a negative correlation between the edges.
An example is the second-order anticorrelation due to data qubit energy relaxation (an energy relaxation event cannot be immediately followed by another relaxation event), which may cause slightly negative $p_{ij}$ in a bit-flip repetition code experiment \cite{mcewen2020}.

Figure \ref{fig:suppl-pij-space-first} shows the same data as Fig.\,\ref{fig:suppl-pij-time-first} but with the different ordering of nodes: here we use the ``space-first'' ordering from Eqn.\,(\ref{space-first}). Then each axis contains $N_{\rm r}+1=31$ blocks corresponding to time rounds (grid lines), while $N_{\rm mq}=10$ points within each block correspond to measure qubits. The S-edges are next to the main diagonal, the T-edges are the diagonal lines separated by 10 pixels from the main diagonal, and the ST edges are on the next diagonal line (11 pixels from the main diagonal). The parallel lines in Fig.\,\ref{fig:suppl-pij-time-first} separated by 20, 30, etc.\ pixels from the main diagonal correspond to 2T, 3T, etc.\ edges. The figure clearly shows that temporal correlations can survive for  over 5 rounds.

{\bf Leakage to state \boldmath$|2\rangle$}.  
We attribute the detection-event correlations lasting for several rounds, as seen in Fig.\
\ref{fig:suppl-pij-space-first}, to the leakage to state $|2\rangle$ in data qubits. The same effect causes the ``dirt'' in Fig.\,\ref{fig:suppl-pij-time-first} close to S and T lines, with the magnitude of the correlations for several edge types shown in the right panel of Fig.\,\ref{fig:suppl-pij-histogram}. Note that measure qubits are reset to $|0\rangle$ at every round, so non-computational states can survive only in data qubits. For a typical qubit energy relaxation time of $T_1\simeq 15 \, \mu$s and the round duration of 960 $\mu$s, we would expect that state $|2\rangle$ should survive on a data qubit for about 8 rounds. Examining Figs.\ \ref{fig:suppl-pij-histogram} and \ref{fig:suppl-pij-space-first}, we see that this estimate is in the right ballpark, but the actual decay of the state $|2\rangle$ can be significantly faster due to hopping of leakage, a subject of ongoing research.

We have found that the amount of leakage is sensitive to minor experimental details. The $p_{ij}$ technique can be used for a fast diagnostic to estimate the level of leakage and to find which qubits suffer a bigger leakage. Specialized experiments have shown \cite{mcewen2020} that a typical probability of state $|2\rangle$ in a data qubit is around $4\times 10^{-3}$. This magnitude is consistent with the values we extract from the $p_{ij}$ analysis. While this analysis is somewhat involved, we note that ST$'$ edges have a somewhat similar (though smaller) $p_{ij}$ values due to leakage. For our phase-flip code experiment, the median value for ST$'$-edge errors is $1.3\times 10^{-3}$, while the biggest value (averaged over rounds) is  $3.3\times 10^{-3}$ for data qubit dq2 (as can be seen from Fig.\,\ref{fig:suppl-pij-time-first}, dq2 has the biggest leakage). So, as a crude proxy for leakage, we can use 
    \begin{align}
    p_{ij}^{\rm ST'-edge \, (leakage)} \alt 3\times 10^{-3}.  
    \end{align}
The 2T edges can also be used to estimate leakage; the biggest 2T-edge value (averaged over rounds) is $3.6\times 10^{-3}$ for measure qubit mq2. (All these values are for the phase-flip code; for a bit-flip code there is an additional contribution from ``odd-even correlations'' due to energy relaxation of data qubits).

Note that during several rounds while a data qubit is in state $|2\rangle$, there is a relatively high probability of detection events at the neighboring measure qubits \cite{mcewen2020}. This leads to a significant correlation between S-edges (and also T-edges), which negatively affects performance of the minimum-weight-matching decoder. This is why leakage is dangerous for quantum error correction even for a relatively low leakage probability.

\begin{figure}
\centering
\includegraphics[width=85mm, trim =3cm 3cm 3cm 3cm,clip=true]{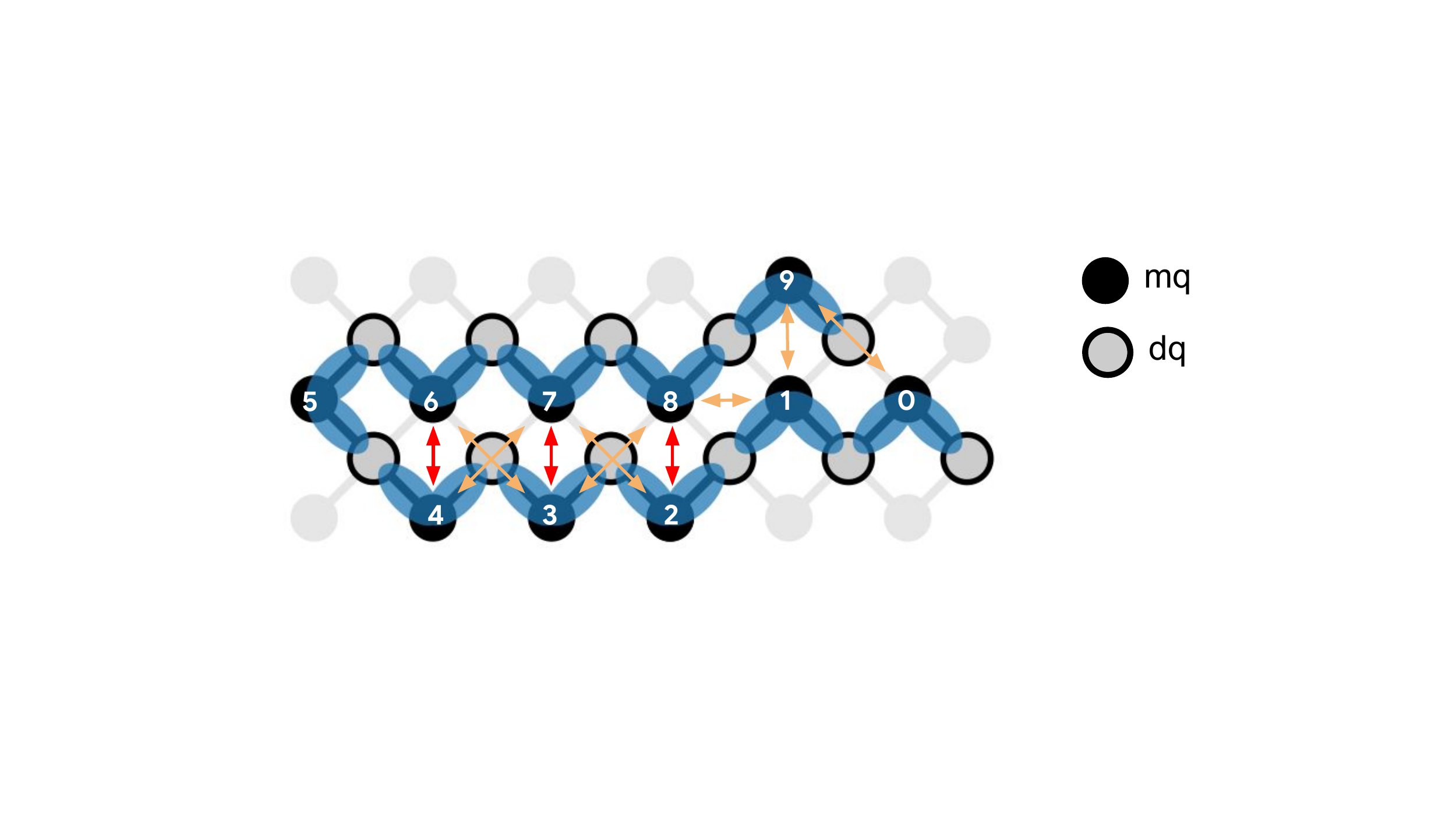}
\includegraphics[width=85mm, trim =2cm 1cm 0cm 1cm,clip=true]{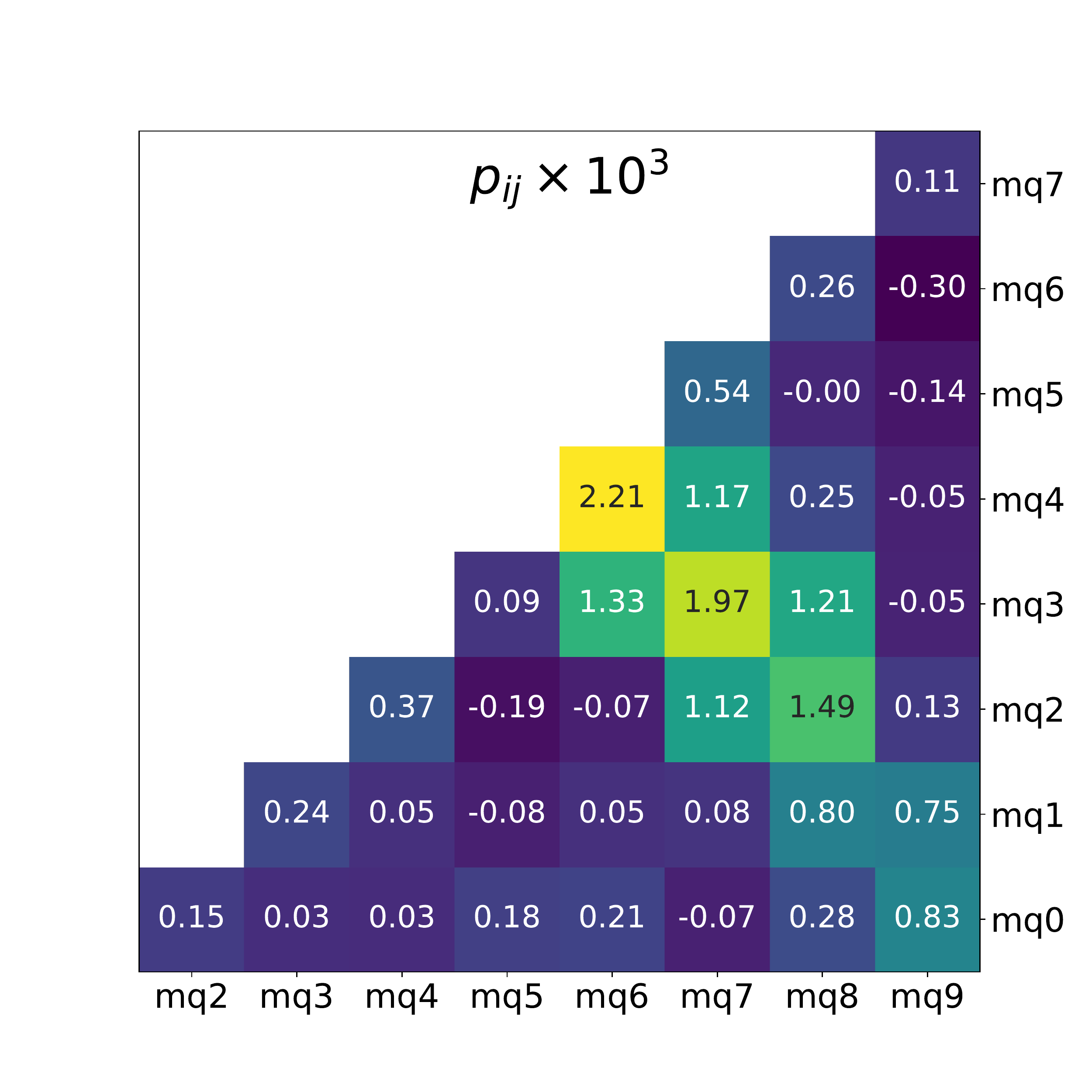}
\caption{ {\bf Crosstalk error probabilities}. Top panel: Layout of 10 measure qubits (black circles with integer labels) and 11 data qubits (gray-filled circles) on the Sycamore device. Arrows indicate the pairs of measure qubits that exhibit stronger (red arrows) and weaker (orange arrows) detection-event correlations  due to crosstalk. Bottom panel: Effective crosstalk probabilities between pairs of measure qubits (except for nearest neighbors). We show the values of $p_{ij}\times 10^3$ for same-round $p_{ij}$ elements averaged over rounds. Cells are colored according to the values: yellow and green indicate a significant crosstalk, blue indicates statistical noise. The biggest crosstalk of $2.2\times 10^{-3}$ is between mq4 and mq6 (leftmost arrow in the top panel).
}
\label{fig:suppl-pij-crosstalk} 
\end{figure}

{\bf Crosstalk features.} Short parallel lines (``scar'' features) in Fig.\,\ref{fig:suppl-pij-time-first} far away from the main diagonal indicate the presence of correlations between detection events at qubits, which are far apart along the 1D line of qubits used in the experiment. However, they are actually close to each other on the Sycamore chip -- see the top panel of Fig.\,\ref{fig:suppl-pij-crosstalk}, which shows 10 pairs of measure qubits (indicated by arrows), for which there are visible scars in Fig.\,\ref{fig:suppl-pij-time-first}. We attribute these scar features to the crosstalk.

The lower panel of Fig.\,\ref{fig:suppl-pij-crosstalk} shows the values of same-round $p_{ij}$ elements averaged over the rounds, for all pairs of measure qubits except nearest neighbors. While most values are within the statistical noise level, the elements corresponding to the scar features are significantly above the noise floor (bigger values are indicated by orange and green cells). We see that the magnitude of the crosstalk correlations is
    \begin{align}
    p_{ij}^{\rm crosstalk} \alt 2\times 10^{-3}.
    \end{align}
For the crosstalk pairs shifted in time by one round we find crudely twice smaller edge probabilities. 

The long-range correlation between detection events caused by crosstalk are dangerous to the code operation because they can effectively reduce the code distance. However, we see that in our device the crosstalk is quite small and, most importantly, local in physical distance on the chip. Therefore, we expect that in the future it will not present a serious problem in a surface code operation.

\FloatBarrier  
\section{Comparison of Edge Weighting Methods for Matching}
\begin{table*}[t]
\caption{\label{table:error}
Error suppression factors ($\Lambda_x$, $\Lambda_z$ for phase and bit flip) and multiplicative constants ($C_x$ and $C_z$) fit to logical error rates vs code distance (Eqn.\,1 of the main text) for the four different edge weighting methods.
}
\begin{ruledtabular}
\begin{tabular}{lcccc}
Weighting method       & $C_x$             & $\lambda_x$      & $C_z$              & $\lambda_z$     \\ \hline
Uniform                & $0.056 \pm 0.005$ & $2.79 \pm 0.056$ & $0.066 \pm 0.007$  & $2.75 \pm 0.06$ \\
Bootstrapping          & $0.068 \pm 0.008$ & $3.18 \pm 0.08$  & $0.078 \pm 0.01$   & $3.01 \pm 0.09$ \\
Correlation ($p_{ij}$) & $0.067 \pm 0.008$ & $3.18 \pm 0.08$  & $0.077 \pm 0.011$  & $3.01 \pm 0.09$ \\
First principles       & $0.067 \pm 0.007$ & $3.17 \pm 0.08$  & $0.0756 \pm 0.011$ & $2.99 \pm 0.09$
\end{tabular}
\end{ruledtabular}
\end{table*}

To decode the error detections obtained in the experiment, we use a minimum weight perfect matching algorithm to determine which physical errors were most likely given the observed directions. 
A key component of this algorithm is the weighting of the edges in the error graph which correspond to the expected correlated probabilities of pairs of nodes. 
The weight of a particular edge ($W$) and the expected probability for that edge ($p$) are related by 

\begin{equation}
W = - \log{p}   
\end{equation}

which satisfies the property that adding the weights of two edges corresponds to multiplying their probabilities. 
We considered four candidate strategies for determining expected edge probabilities and weights:

\begin{enumerate}
    \item \textbf{Uniform weighting} - assume that all edges in the matching graph are equally likely
    \item \textbf{Bootstrapping} - Run matching on a training dataset with uniform weights, then for a given edge, count the number of times it was matched and divide by the number of total experiments to compute the expected probability for future matches.
    \item \textbf{Node correlations ($p_{ij}$)} - Use the node correlation technique described in Section \ref{section:pij} to determine the correlated probabilities for edges from a training dataset.
    \item \textbf{First principles} - From the measured gate, measurement, and reset error probabilities, compute the edge probabilities by propagating possible errors through the circuit.
\end{enumerate}

For methods 2 and 3, we use the data at 50 rounds to determine the matching weights for all other datasets. While these methods can in general produce a unique weight for each edge in the 50 round graph, we average together all rounds so that the edge weights used during matching are uniform in time. Phase flip and bit flip edge weights, as well as weights for each of the smaller subsampled codes, are determined separately.

In Tab.\,\ref{table:error}, we show the fitted values of $\Lambda$ using the different weighting methods, for both the bit and phase flip codes. 
To within the uncertainty from fitting, we find that methods 2, 3, 4 all give the same result for $\Lambda_x$ and $\Lambda_z$, while uniform weighting reduces $\Lambda_x$ to 2.7 and $\Lambda_z$ to 2.5. 
The primary effect of the more sophisticated weighting methods is to increase the weights of spacetime edges relative to spacelike and timelike edges.

\section{Dynamical decoupling of data qubits}
\begin{figure}
\includegraphics[width=85mm]{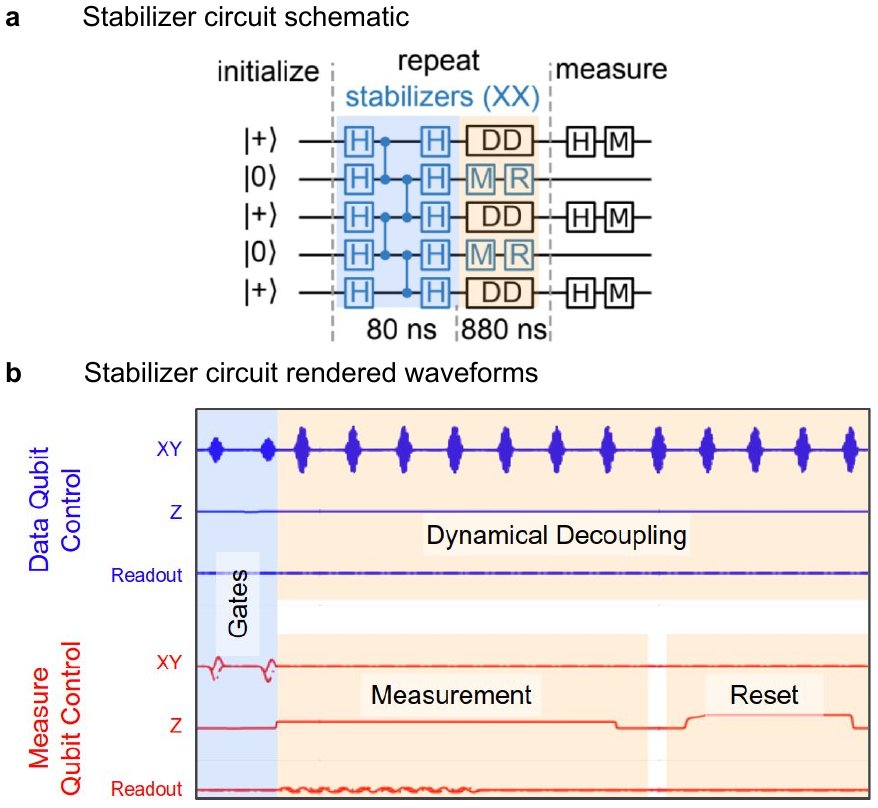}
\caption{\label{fig:sx1}\textbf{Stabilizer Circuit.} \textbf{a,} Circuit schematic representation of the stabilizer circuit. Layers of single qubit and two qubit gates highlighted in blue. Measurement, reset, and dynamical decoupling operations highlighted in yellow to correspond to the waveforms in \textbf{b,} Rendered waveforms to show that the majority of the time spent during the stabilizer is during the measurement and reset operations. Lines represent microwave control (XY), flux control (Z), and readout for the stabilizer circuit for one data qubit (blue) and one measure qubit (red).}
\end{figure}

\begin{figure}
\includegraphics{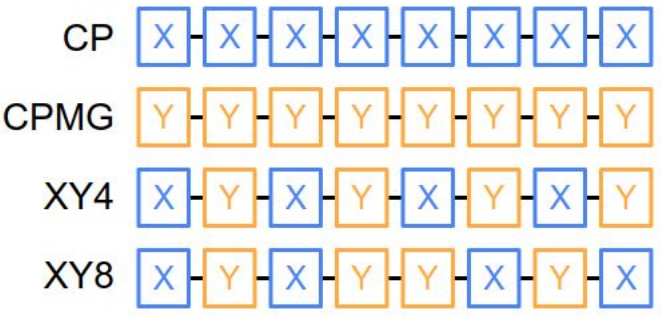}
\caption{\label{fig:sx2}\textbf{Dynamical Decoupling Sequences.} The four multi-pulse sequences used during the measurement and reset portions of the stabilizer circuit. Each sequence has the same total idle time and executes the same number of gates. The distinction between these four sequences during the execution of the circuit is only the phase of the microwave pulses, a technique used to compensate for cumulative pulse errors.}
\end{figure}

\begin{figure*} [t]
\includegraphics[width=176mm]{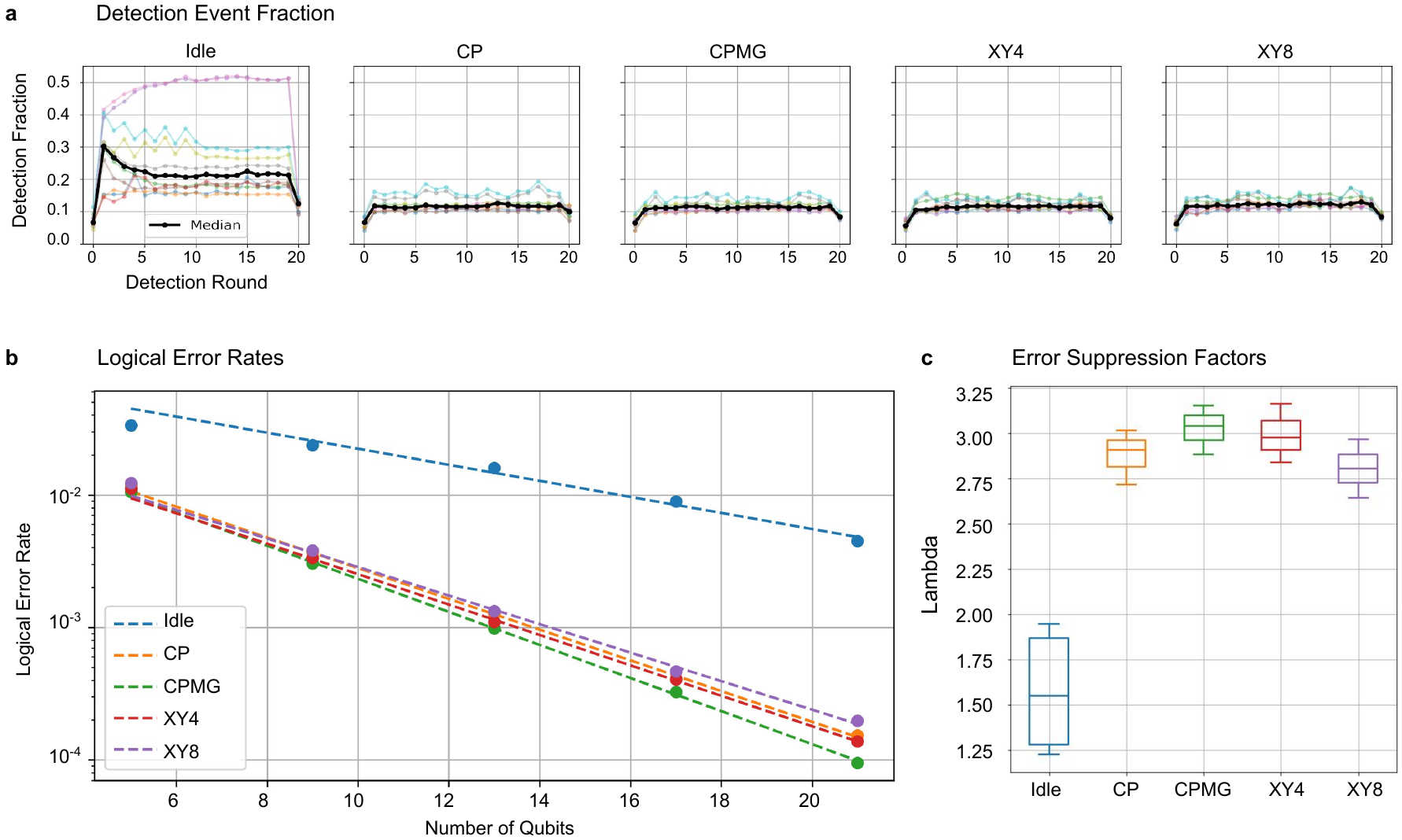}
\caption{\label{fig:sx3}\textbf{Benchmarking phase flip performance with and without dynamical decoupling.} b Detection event fractions vs qubit and round for each of the data qubit Idle, CP, CPMG, XY4, and XY8 operations during measure qubit readout and reset. Median detection event fraction by round plotted in black. \textbf{b,} Logical error rate vs number of qubits, showing exponential suppression of error rate in all cases. \textbf{c,} Boxplot of extracted error suppression factors ($\Lambda$) from fits like those shown in \textbf{b,} for five iterations of the experiment for each decoupling scheme. Overall, we see an ~1.7x increase in $\Lambda$ for all decoupling schemes. The performance between the various decoupling schemes is comparable.}
\end{figure*}

The measurement and reset operations take 880\,ns to complete and account for approximately 92\% of the time spent for the duration of the phase flip code (see Fig.\,\ref{fig:sx1}). 
Leaving data qubits to idle during these operations, we undergo energy relaxation processes in addition to dephasing processes, accounting for a large portion of the total error budget. 
The process of measurement and reset on the measure qubits introduces additional avenues of error including measurement-induced dephasing from photon crosstalk between readout resonators \cite{szombati2020quantum}, as well as frequency detuning errors incurred from any flux crosstalk between qubits. 
While energy relaxation is irreversible and cannot be mitigated here, dephasing can be mitigated using dynamical-decoupling techniques. 
We employ multi-pulse sequences developed within the field of NMR which have been shown to mitigate low-frequency noise in superconducting qubits \cite{bylander2011noise}: Carr-Purcell (CP) \cite{carr1954effects}, Car-Purcell-Meiboom-Gill (CPMG) \cite{meiboom1958modified}, XY4, and XY8 \cite{gullion1990new}.  

With independent phase coherence measurements, we verified that we were able to effectively decouple the qubits from the noise sources listed above.
Using CPMG, we verified independently via phase coherence measurements with and without adversarial readout tones, as well as with and without large frequency excursions on neighboring qubits, that we are able to effectively decouple away the intrinsic low-frequency noise, measurement-induced dephasing on the data qubits caused by crosstalk from measure, as well as any flux crosstalk effects. 
We then evaluated the performance of each dynamical decoupling protocol within the context of the repetition code. For all of the decoupling sequences, we fix the time between pulses such that every sequence has the same total idle time and executes the same number of gates (see Fig.\,\ref{fig:sx2}). 
The fixed idle time was set such that each sequence performed eight gates. Using decoupling, we see an ${\sim}1.7 \times$ increase in the error suppression factor, $\Lambda$ (Fig.\,\ref{fig:sx3}). 
To compare the performance of the different decoupling schemes, the experiment was run and analyzed a total of five times for each of the schemes (Idle, CP, CPMG, XY4, and XY8). 
The performance between schemes was comparable with the CPMG and XY4 sequences slightly outperforming the CP and XY8 sequences.

\FloatBarrier  
\section{Qubit Frequency Optimization}

\begin{figure*} [t]
\includegraphics[width=183mm]{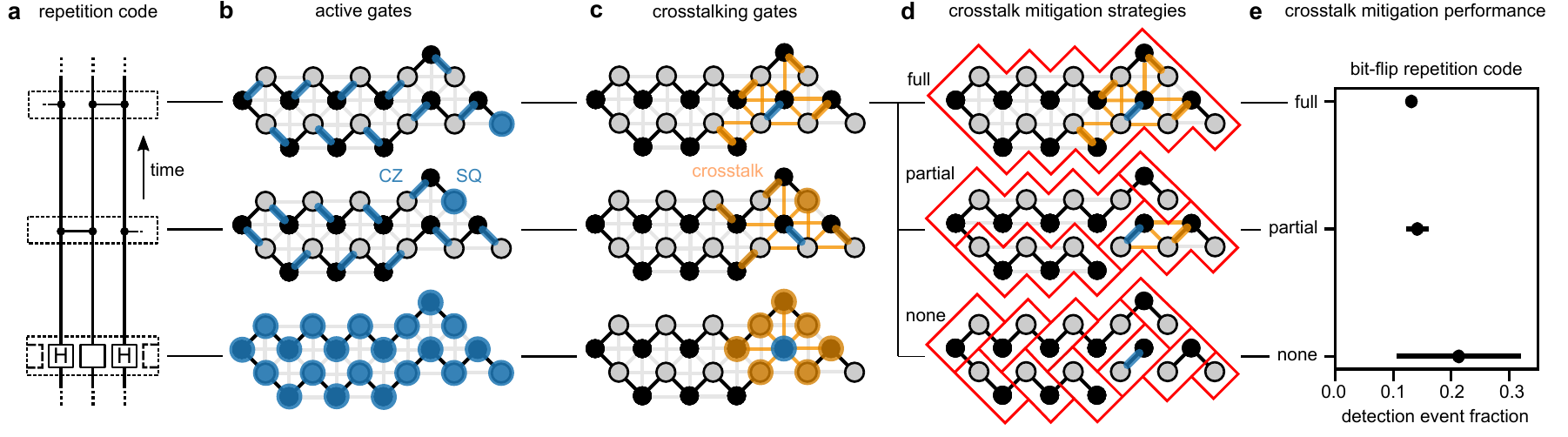}
\caption{\label{fig:snake} \textbf{Qubit-crosstalk mitigation.} \textbf{a,} The repetition code, with three distinct temporal slices indicated by dashed boxes. The empty boxes in the lowest temporal slice are either H or I depending on whether we run the bit- or phase-flip code. \textbf{b,} Simultaneously active SQ (H or I) and CZ gates (blue nodes and edges, respectively) at each temporal slice. The geometry of active gates is determined by the repetition code circuit and its mapping onto our processor. Simultaneously active gates can crosstalk due to parasitic interactions between NN and NNN qubits. \textbf{c,} Crosstalking SQ and CZ gates (orange nodes and edges, respectively) for one active SQ or CZ (blue nodes and edges, respectively) gate at each temporal slice. We mitigate crosstalk and other error mechanisms by constructing and optimizing an error model with respect to gate-frequencies. \textbf{d,} Three crosstalk mitigation strategies illustrated for one active CZ gate in the upper temporal slice in \textbf{a} - \textbf{c}. The strategies are labelled “full”, “partial”, and “none”, according to the degree of expected crosstalk protection. Each strategy can be characterized by domains (red) in which crosstalk is penalized. \textbf{e,} Bit-flip repetition code benchmarks for each mitigation strategy. The points and error bars represent the DEF median and standard-deviation, respectively. By increasing the mitigation strength from “none” to “full”, the DEF median and standard-deviation are reduced by 38\% and 91\%, respectively.}
\end{figure*}

Our processor employs frequency-tunable qubits \cite{arute2019quantum}. Quantum logic gates are executed at two distinct types of frequencies: idle and interaction frequencies, which are collectively referred to as gate frequencies. Qubits idle and execute single-qubit gates at their respective idle frequencies. Neighboring qubit-pairs execute CZ gates at their respective interaction frequencies. All gate frequencies are explicitly or implicitly interdependent due to engineered interactions and/or crosstalk according to the repetition code circuit and its mapping onto our processor. Since many error mechanisms are frequency dependent, we can mitigate errors by constructing and optimizing an error model with respect to gate frequencies. 

To construct an error model, we combine error contributions from Z pulse-distortion, relaxation, dephasing, and qubit crosstalk. The Z pulse-distortion model penalizes CZ gates for large frequency excursions. The relaxation and dephasing models penalize SQ and CZ gates for approaching relaxation and dephasing hotspots, while incorporating coupler physics, qubit hybridization, state-dependent transitions, and hardware-accurate frequency trajectories. Finally, the qubit-crosstalk model penalizes for frequency collisions between nearest-neighbor (NN) and diagonal next-nearest-neighbor (NNN) qubits, while incorporating qubit hybridization and the mapping of the repetition code circuit onto our processor. These constituent models are determined via theory and/or experiment, consolidated, and then trained to be predictive of experimentally measured error benchmarks via machine learning.

To determine a frequency configuration that mitigates error, we optimize the error model with respect to gate frequencies. Optimization is complex since the error model spans 41 frequency variables, is non-convex, and time-dependent \cite{klimov2018fluctuations}. Furthermore, since each frequency variable is constrained to ${\sim}10^2$ values by the control hardware and qubit-circuit parameters, the optimization search space is ${\sim}2^{272}$, which significantly exceeds the Hilbert-space dimension $2^{21}$. Given the optimization complexity, exhaustive search is intractable and global optimization is too slow and inefficient. To quickly and efficiently find locally optimal gate-frequency configurations and maintain them in the presence of drift, we use our Snake optimizer \cite{klimov2020snake}. 

To illustrate the performance of our error mitigation strategy, we conduct a qubit-crosstalk mitigation experiment (see Fig.\,\ref{fig:snake}). In this experiment, we first optimize our processor employing one of three qubit-crosstalk mitigation strategies. We then calibrate the processor and run the bit-flip repetition code. The three mitigation strategies are labelled “none”, “partial”, and “full”, according to the expected degree of crosstalk protection. In the “none” strategy, we do not penalize for crosstalk. In the “partial” strategy, we penalize for crosstalk according to the cross-entropy benchmarking (XEB) circuit \cite{arute2019quantum}, which we often use in calibration. Although XEB and the repetition code have different circuits and serve different purposes, their respective circuits have similar gate patterns (see Fig. S25 of Ref.\,\cite{arute2019quantum}). Because of this similarity, penalizing for crosstalk according to XEB should also offer partial crosstalk protection for the repetition code. Finally, in the “full” strategy, we penalize for crosstalk according to the repetition code circuit that we run.  

To quantify the efficacy of the three mitigation strategies, we inspect bit-flip repetition-code detection event fractions (DEF). We see that by increasing the degree of crosstalk mitigation from “none” to “partial” to “full”, the median DEF is reduced by 33\% and 7\%, respectively. Furthermore, the DEF standard-deviation is reduced by 82\% and 51\%, respectively. In total, this amounts to a 38\% reduction in median DEF and a 91\% reduction in the DEF standard-deviation, representing a significant performance boost. We delegate error mitigation data for other error mechanisms to a future publication. 

\FloatBarrier  
\section{Overview of error correction experiments}
In Table \ref{table:experiments}, we list experimental implementations of quantum error correction as a reference.

\begin{table*}[t]
\caption{\label{table:experiments}Various error correction and error detection experiments.
    Experiments using ``classical" codes (i.e. codes that only detect one type of error e.g. only phase flips or only bit flips) use classical $[n,k,d]$ code notation instead of quantum $[[n,k,d]]$ code notation.
    Entries with an N/A are experiments related to embedding error correction into the physical qubits as opposed to layering the error correction on top of the physical qubits.
    Note that there is, as of yet, no experiment exploring a range of rounds and a range of code distances using a non-classical code.}
\begin{ruledtabular}
\begin{tabular}{cccccccccc}
Paper & Year & Code name & [[\#data,\#logical,distance]] & Physical qubits & Rounds & Physical qubit type \\
\hline
\cite{cory1998experimental} & 1998 & Repetition Code & [3,1,3] & 3 & single shot & NMR \\
\cite{knill2001benchmarking} & 2001 & Perfect Code & [[5,1,3]] & 5 & single shot & NMR \\
\cite{schindler2011experimental} & 2011 & Repetition Code & [3,1,3] & 3 & 3 & Ion trap \\
\cite{moussa2011demonstration} & 2011 & Repetition Code & [3,1,3] & 3 & 2 & NMR \\
\cite{zhang2011experimental} & 2011 & Repetition Code & [3,1,3] & 3 & single shot & NMR \\
\cite{reed2012realization} & 2012 & Repetition Code & [3,1,3] & 3 & single shot & Superconducting \\
\cite{zhang2012experimental} & 2012 & Perfect Code & [[5,1,3]] & 5 & single shot & NMR \\
\cite{bell2014experimental} & 2014 & Surface Code & [[4,1,2]] & 4 & single shot & Photons \\
\cite{kelly2015state} & 2014 & Repetition Code & [3,1,3]-[5,1,5] & 9 & 8 & Superconducting \\
\cite{nigg2014quantum} & 2014 & Color Code & [[7,1,3]] & 7 & single shot & Ion trap \\
\cite{waldherr2014quantum} & 2014 & Repetition Code & [[3,1,3]] & 4 & single shot & NV center \\
\cite{riste2015detecting} & 2015 & Repetition Code & [3,1,3] & 5 & single shot & Superconducting \\
\cite{corcoles2015demonstration} & 2015 & Bell State & [[2,0,2]] & 4 & single shot & Superconducting \\
\cite{cramer2016repeated} & 2016 & Repetition Code & [3,1,3] & 4 & 1-3 & Superconducting \\
\cite{ofek2016extending} & 2016 & Cat States & N/A & 1 & 1-6 & 3D cavity \\
\cite{takita2017experimental} & 2017 & Color Code & [[4,2,2]] & 5 & single shot & Superconducting \\
\cite{linke2017fault} & 2017 & Color Code & [[4,2,2]] & 5 & single shot & Ion trap \\
\cite{wootton2018repetition} & 2018 & Repetition Code & [3,1,3]-[8,1,8] & 15 & single shot & Superconducting \\
\cite{andersen2019entanglement} & 2019 & Bell State & [[2,0,2]] & 3 & 1-12 & Superconducting \\
\cite{gong2019experimental} & 2019 & Perfect Code & [[5,1,3]] & 5 & single shot & Superconducting \\
\cite{hu2019quantum} & 2019 & Binomial Bosonic States & N/A & 1 & 1-19 & 3D cavity \\
\cite{wootton2020benchmarking} & 2020 & Repetition Code & [3,1,3]-[22,1,22] & 5-43 & single shot & Superconducting \\
\cite{andersen2020} & 2020 & Surface Code & [[4,1,2]] & 7 & 1-11 & Superconducting \\
\cite{bultink2020protecting} & 2020 & Bell State & [[2,0,2]] & 3 & 1-26 & Superconducting \\
\cite{egan2020fault} & 2020 & Bacon-Shor Code & [[9,1,3]] & 15 & single shot & Ion trap \\
\cite{luo2020quantum} & 2020 & Bacon-Shor Code & [[9,1,3]] & 11 & single shot & Photons \\
\cite{campagne2020quantum}& 2020 & GKP States & N/A & 1 & 1-200 & 3D cavity \\
This work & 2020 & Repetition Code & [3,1,3]-[11,1,11], [[4,1,2]] & 5-21 & 1-50 & Superconducting \\
\end{tabular}
\end{ruledtabular}
\end{table*}

\FloatBarrier 

\end{document}